\documentstyle[preprint,aps]{revtex}
\begin{document}
\draft
\title{Stochastic Processes and Diffusion on Spaces with Local Anisotropy}
 \author{Sergiu I.\ Vacaru}
\address{Institute of Applied Physics, Academy of Sciences of Moldova,\\
5 Academy str., Chi\c{s}in\u{a}u 277028, Republic of Moldova}
\date{\today}
\maketitle
\begin{abstract}
The purpose of this work is to extend the formalism of stochastic calculus
to the case of spaces with local anisotropy (modeled as vector bundles with
 compatible nonlinear and distinguished connections and metric structures and
 containing as particular cases different variants of Kaluza--Klein and
generalized Lagrange and Finsler spaces). We shall examine nondegenerate
diffusions on the mentioned spaces and theirs horizontal lifts.

{\bf \copyright S.I. Vacaru}
\end{abstract}
\narrowtext

\section{Introduction}

Probability theorists, physicists, biologists and financiers are already
familiar with classical and quantum statistical and geometric methods
applied in various branches of science and economy [1-6]. We note that
modeling of diffusion processes in nonhomogerneous media and formulation of
nonlinear thermodynamics in physics, or of dynamics of evolution of species
in biology, requires a more extended geometrical background than that used
in the theory of stochastic differential equations and diffusion processes
on Riemann and Lorentz manifolds [7-10].

Our aim is to formulate the theory of diffusion processes on spaces with
local anisotropy. As a model of such spaces we choose vector bundles on
space-times provided with nonlinear and distinguished connections and metric
structures [11,12]. Transferring our considerations on tangent bundles we
shall formulate the theory of stochastic differential equations on
generalized Lagrange spaces which contain as particular cases Lagrange and
Finsler spaces [13-17].,

The plan of the presentation is as follow: In Section II we give a brief
summary of the geometry of locally anisotropic spaces. Section III is
dedicated to the formulation of the theory of stochastic differential
equations in vector bundle spaces. This section also concerns the basic
aspects of stochastic calculus in curved spaces. In Section IV the heat
equations in bundle spaces are analyzed. The nondegenerate diffusion on
spaces with local anisotropy is defined in Section V. We shall generalize in
Section VI the results of Section IV to the case of heat equations for
distinguished tensor fields in vector bundles with (or not) boundary
conditions. Section VII contains concluding remarks and a discussion of the
obtained results.\ We present a brief introduction into the theory of
stochastic differential equations and diffusion processes on Euclidean
spaces in the Appendix.

\section{Nonlinear and Distinguished Connections and Metric Structures in
Vector Bundles}

As a preliminary to the discussion of stochastic calculus on locally
anisotropic spaces we summarize some modern methods of differential geometry
of vector bundles, generalized Lagrange and Finsler spaces. This Section
serves the twofold purpose of establishing the geometrical background and
deriving equations which will be used in the next two Sections. In general
lines we follow conventions from [11,12].

Let ${\cal E} = (E,\pi ,F,Gr,M)$ be a locally trivial vector bundle,
v-bundle, where \thinspace $F={\cal R}^m$ is a real $m$ dimensional vector
space, ${\cal R\ }$denotes the real number field , $Gr=GL(m,{\cal R)\,}$is
the group of automorphisms of ${\cal R}^m$, base space $M$ is a
differentiable (class $C^\infty $) manifold of dimension $n,$ $\dim M=n,\pi
:E\rightarrow $ $M$ is a surjective map and the differentiable manifold $%
E,\dim E=n+m=q,$ is called the total space of v-bundle ${\cal E\ }$ . We
locally parametrize space ${\cal E\ \,}$ by coordinates $u^\alpha
=(x^i,y^a),\,$ where $x^i$ and $y^{a\,}$ are, correspondingly, local
coordinates on $M$ and $F.\,\,$ We shall use Greek indices for cumulative
coordinates on ${\cal E\ }$ , Latin indices $i,j,k,...=0,1,...,n-1\,$ for
coordinates on $M\,$ and $a,b,c,...=1,2,...,m$ for coordinates on $F.$
Changings of coordinates $(x^k,y^a)\rightarrow (x^{k^{\prime }},y^{a^{\prime
}})$ on ${\cal E\ }$ are written as

$$
x^{k^{\prime }}=x^{k^{\prime }}(x^k),y^{a^{\prime }}=M_a^{a^{\prime
}}(x)y^a,
$$
\begin{equation}
rank\left( \frac{\partial x^{k^{\prime }}}{\partial x^k}\right)
=n,M_a^{a^{\prime }}(x)\in Gr.\eqnum{2.1}
\end{equation}

A nonlinear connection, N-connection, in v-bundle ${\cal E}$ is defined as a
global decomposition into horizontal ${\cal HE\,}$ and vertical ${\cal VE\,}$
subbundles of the tangent bundle:%
\begin{equation}
{\cal TE=HE\oplus VE}.\eqnum{2.2}
\end{equation}
N-connection defines a corresponding covariant derivation in ${\cal E:}$
\begin{equation}
\nabla _YA=Y^i\{\frac{\partial A^a}{\partial x^i}+N_i^a\left( s,A\right)
\}s_a,\eqnum{2.3}
\end{equation}
where $s_a$ are local linearly independent sections of ${\cal E,\ }A=A^as_a$
and $Y=Y^is_i$ is a vector field decomposed on local basis $s_i$ on $M.\,$
Differentiable functions $N_i^a\,\,$ from (2.3), written as functions on $%
x^i $ and $y^a,$ i.e. as $N_i^a\left( x,y\right) ,$ are called the coefficients
of N-connection.

In v-bundle ${\cal E\,}$ we can define a local frame (basis) adapted to a
given N-connection:%
\begin{equation}
\delta _\alpha =\frac \delta {\delta u^\alpha }=(\delta _i=\frac \delta
{\delta x^i}=\frac \partial {\partial x^i}-N_i^a(x,y)\frac \partial
{\partial y^a},\delta _a=\frac \delta {\delta y^a}=\partial _a=\frac
\partial {\partial y^a}).\eqnum{2.4}
\end{equation}
The dual basis is written as%
\begin{equation}
\delta ^\alpha =\delta u^\alpha =(\delta ^i=\delta x^i=d^i=dx^i,\delta
^a=\delta y^a=dy^a+N_i^a(x,y)dx^i).\eqnum{2.5}
\end{equation}

By using adapted bases one introduces the algebra of tensorial distinguished
fields (d-fields or d-tensors) on
 ${\cal E}, {\cal T} =  \oplus {\cal T}_{qs}^{pr},$ which
is equivalent to the tensor algebra on v-bundle ${\cal E}_d$ defined as
$$
\pi _d:{\cal HE\oplus VE\rightarrow E.\ }
$$
An element $t\in {\cal T}_{qs}^{pr},\ $ d-tensor field of type $\left(
\begin{array}{cc}
p & r \\
q & s
\end{array}\right)
,$ can be written in local form as
$$
t=t_{j_1...j_qb_1...b_s}^{i_1...i_pa_1...a_r}(x,y)\delta _{i_1}\otimes
...\otimes \delta _{i_p}\otimes d^{j_1}\otimes ...\otimes d^{j_q}\otimes
$$
$$
\partial _{a_1}\otimes ...\otimes \partial _{a_r}\otimes \delta
^{b_1}\otimes ...\otimes \delta ^{b_s}.
$$

In addition to d-tensors we can consider d-objects with various group and
coordinate transforms adapted to a global splitting (2.2). For our further
considerations the linear d-connection structure, defined as a linear
connection $D$ in ${\cal E\ }$ conserving under parallelism the Whitney sum $%
{\cal HE\oplus VE\ \,\,}$ associated to a fixed N-connection in ${\cal E\ }$%
will play an important role. Components $\Gamma _{\beta \gamma }^\alpha $ of
d-connection $D$ are introduced as%
$$
D_\gamma \delta _\beta =D_{\delta _\gamma }\delta _\beta =\Gamma _{\beta
\gamma }^\alpha \delta _\alpha .
$$

Torsion $T_{\beta \gamma }^\alpha $ and curvature $R_{\beta \gamma \delta
}^\alpha $ of d-connection $\Gamma _{\beta \gamma }^\alpha $ are defined in
a standard manner:%
$$
T(\delta _\gamma ,\delta _\beta )=T_{\beta \gamma }^\alpha \delta _\alpha ,
$$
where%
\begin{equation}
T_{\beta \gamma }^\alpha =\Gamma _{\beta \gamma }^\alpha -\Gamma _{\gamma
\beta }^\alpha +w_{\beta \gamma}^{\alpha}, \eqnum{2.6}
\end{equation}
and
$$
R(\delta _\tau ,\delta _\gamma ,\delta _\beta )=R_{\beta \gamma \tau
}^\alpha \delta _\alpha ,
$$
where
\begin{equation}
R_{\beta \gamma \tau }^\alpha =\delta _\tau \Gamma _{\beta \gamma }^\alpha
-\delta _\gamma \Gamma _{\beta \tau }^\alpha +\Gamma _{\beta \gamma
}^\varphi \Gamma _{\varphi \tau }^\alpha -\Gamma _{\beta \tau }^\varphi
\Gamma _{\varphi \gamma }^\alpha +\Gamma _{\beta \varphi }^\alpha w_{\gamma
\tau }^\varphi .\eqnum{2.7}
\end{equation}
In formulas (2.6) and (2.7) we have used nonholonomy coefficients $w_{\beta
\gamma }^\alpha $ of the adapted frames (2.3)%
\begin{equation}
[\delta _\alpha ,\delta _\beta ]=\delta _\alpha \delta _\beta -\delta _\beta
\delta _\alpha =w_{\alpha \beta }^\gamma \delta _{\beta .}\eqnum{2.8}
\end{equation}

Another important geometric structure on ${\cal E\ }$ which will be
considered is the metric structure on $G,\,$ defined as a nondegenerate,
second order, with constant signature (in this work positive), tensor field $%
G_{\alpha \beta }$ on ${\cal E}.$ We can associate a map
$$
G(u):T_u{\cal E}\times T_u{\cal E}\rightarrow {\cal R}
$$
to a metric%
$$
G=G_{\alpha \beta }du^\alpha \otimes du^\beta
$$
parametrized by a sym\-met\-ric, $rankG=q,$ matrix $$\left(
\begin{array}{cc}
G_{ij} & G_{ia} \\
G_{aj} & G_{ab}
\end{array}\right)
,$$ where $G_{ij}=G\left( \partial _i,\partial _j\right) ,G_{ia}=G\left(
\partial _i,\partial _a\right) ,G_{ab}=G\left( \partial _a,\partial
_b\right) .$

We shall be interested by such a concordance of N-connection and metric
structures when $G\left( \delta _i,\partial _a\right) =0,$ or, equivalently,%
\begin{equation}
N_i^a(x,y)=G_{ib}(x,y)G^{ba}(x,y),\eqnum{2.9}
\end{equation}
where matrix $G^{ba}$ is inverse to matrix $G_{ab}.$ In this case metric $G$
on {\cal E} is defined by two independent d-tensors, $g_{ij}$ of type $\left(
\begin{array}{cc}
2 & 0 \\
0 & 0
\end{array}\right)
$and $h_{ab}$ of type $\left(
\begin{array}{cc}
0 & 0 \\
0 & 2
\end{array}\right)
,\,$ and can be written with respect to bases (2.4) and (2.5) as%
\begin{equation}
G=g_{ij}(x,y)dx^i\otimes dx^j+h_{ab}(x,y)\delta y^a\otimes \delta y^b
\eqnum{2.10}
\end{equation}

We shall call a metric $G$ compatible with a d-connection $D$ if conditions%
\begin{equation}
D_\alpha G_{\beta \gamma }=0 \eqnum{2.11}
\end{equation}
are satisfied.

{\bf Definition 1.}{\it \ A vector bundle } ${\cal E}$ {\it on base }$M$ {\it %
provided with compatible N-connection, d-connection and metric structures
(when conditions (2.9) and (2.11) are satisfied) is called a space with
local anisotropy, (la-space) and denoted as }${\cal E_N.}$

{\bf Remarks.}

1. For the case when instead of v-bundle ${\cal E}$ the tangent bundle $TM\,$
is considered the conditions (2.9), (2.11) and the requirement of
compatibility of compatibility of N-connections with the almost Hermitian
structure on $TM\,$ lead to the equality [11,12] (see metric (2.10) on $TM\,$
)

$$
h_{ij}(x,y)=g_{ij}(x,y).
$$
The metric field $g_{ij}(x,y)$ is the most general form of metric structure
with local anisotropy, considered in generalized Lagrange geometry.

2. Metrics on a Lagrange space $(M,{\cal L})$ can be considered as a
particular case of metrics of type (2.10) on $TM$ for which there is a
differentiable function ${\cal L:}TM\rightarrow {\cal R}$ with the property
that d-tensor%
\begin{equation}
g_{ij}(x,y)=\frac 12\frac{\partial ^2{\cal \ L}}{\partial y^i\partial y^j}%
\eqnum{2.12}
\end{equation}
is nondegenerate.

3. We obtain a model of Finsler space $(M,L)$ if ${\cal L=\ }L^2,$ where $L$
is a Finsler metric [11-12].

We emphasize that in this work all geometric constructions and results on
stochastic calculus will be formulated for the general case of la-spaces.

On a la-space ${\cal E_N}$ we can consider arbitrary compatible with metric $%
G$ d-connections $\Gamma _{\beta \gamma }^\alpha ,$ which are analogous of
the affine connections on locally isotropic spaces (with or not torsion). On
${\cal E_N}$ it is defined the canonical d-structure$\overrightarrow{\Gamma }%
_{\beta \gamma }^\alpha $ with coefficients generated by components of
metric and N-connection%
$$
\overrightarrow{\Gamma }_{jk}^i=L_{jk}^i,\overrightarrow{\Gamma }%
_{ja}^i=C_{ja}^i,\overrightarrow{\Gamma }_{aj}^i=0,\overrightarrow{\Gamma }%
_{ab}^i=0,
$$
\begin{equation}
\overrightarrow{\Gamma }_{jk}^a=0,\overrightarrow{\Gamma }_{jb}^a=0,
\overrightarrow{\Gamma }_{bk}^a=L_{bk}^a,\overrightarrow{\Gamma }%
_{bc}^a=C_{bc,}^a\eqnum{2.13}
\end{equation}
where%
$$
L_{jk}^i=\frac 12g^{ip}(\delta _kg_{pi}+\delta _jg_{pk}-\delta _pg_{jk}),
$$
$$
L_{bi}^a=\partial _bN_i^a+\frac 12h^{ac}(\delta _ih_{bc}-\partial
_bN_i^dh_{dc}-\partial _cN_i^dh_{db}),
$$
$$
C_{jc}^i=\frac 12g^{ik}\partial _bg_{jk},
$$
$$
C_{bc}^a=\frac 12h^{ad}(\partial _ch_{db}+\partial _bh_{dc}-\partial
_dh_{bc}).
$$
In formulas (2.13) we have used matrices $g^{ij}$ and $h^{ab}\,$which are
respectively inverse to matrices $g_{ij}$ and $h_{ab}.$

We also present the explicit formulas for unholonomy coefficients $w_{\beta
\gamma }^\alpha \,\,$of the adapted frame basis (2.4):
$$
w_{ij}^k=0,w_{aj}^k=0,w_{ia}^k=0,w_{ab}^k=0,w_{ab}^c=0,
$$
\begin{equation}
w_{ij}^a=R_{ij}^{a\,},w_{ai}^b=-\partial _aN_i^b,w_{ia}^b=\partial
_aN_i^b.\eqnum{2.14}
\end{equation}
Putting (2.13) and (2.14) into, correspondingly, (2.6) and (2.7) we can
computer the components of canonical torsion $\overrightarrow{T}_{\beta
\gamma }^\alpha $ and curvature $\overrightarrow{R}_{\beta \gamma \delta
}^\alpha $ with respect to the locally adapted bases (2.4) and (2.5) (see
[10,11]).

Really, on every la-space ${\cal E_N}$ a linear multiconnection d-structure
is defined. We can consider at the same time some ways of local transports
of d-tensors by using, for instance, an arbitrary d-connection $\Gamma
_{\beta \gamma }^\alpha ,$ the canonical one $\overrightarrow{\Gamma }%
_{\beta \gamma }^\alpha ,$ or the so-called Christoffel d-symbols defined as%
\begin{equation}
\{\frac \alpha {\beta \gamma }\}=\frac 12G^{\alpha \tau }(\delta _\gamma
G_{\beta \tau }+\delta _\beta G_{\gamma \tau }-\delta _\tau G_{\beta \gamma
}).\eqnum{2.15}
\end{equation}
Every compatible with metric d-connection $\Gamma _{\beta \gamma }^\alpha $
can be characterized by a corresponding deformation d-tensor with respect,
for simplicity, to $\{\frac \alpha {\beta \gamma }\}:$%
\begin{equation}
P_{\beta \gamma }^\alpha =\Gamma _{\beta \gamma }^\alpha -\{\frac \alpha
{\beta \gamma }\}\eqnum{2.16}
\end{equation}
(the deformation of the canonical d-connection is written as
\begin{equation}
\overrightarrow{P}_{\beta \gamma }^\alpha =\overrightarrow{\Gamma }_{\beta
\gamma }^\alpha -\{\frac \alpha {\beta \gamma }\}).\eqnum{2.17}
\end{equation}
Perhaps, it is more convenient to consider physical models and geometric
constructions with respect to the torsionless d-connection $\{\frac \alpha
{\beta \gamma }\}.\,$The more general ones will be obtained by using
deformations of connections of type (2.16). But sometimes it is possible to
write out d-covariant equations on ${\cal E_N}$ , having significance, by
changing respectively components $\{\frac \alpha {\beta \gamma }\}$ on $%
\Gamma _{\beta \gamma }^\alpha $ . This holds for definition of stochastic
differential equations on la-spaces (see, in particular, [1,33] on diffusion
on Finsler spaces) and in this paper we shall use the last way.

Let suppose that ${\cal E_N}$ is locally trivial and $\sigma $-compact. In
this case ${\cal E_N}$ is a paracompact manifold and has a countable open
base. We denote as $F({\cal E_N\ )}$ the set of all real $C^\infty $%
-functions on ${\cal E_N}$ and as $F_0\left( {\cal E_N}\right) $ the
subclass of $F\left( {\cal E_N}\right) $ consisting from functions with
compact carriers. $F_0\left( {\cal E_N}\right) $ and $F\left( {\cal E_N}%
\right) $ are algebras on the field of real numbers ${\cal R\ }$ with usual
operations $f+q,fq$ and $\lambda f(f,q\in F({\cal E_N\ )}$ or $F_0\left(
{\cal E_N}\right) ,\lambda \in {\cal R).}$

Vector fields on ${\cal E_N}$ are defined as maps%
$$
V:u\in {\cal E_N\ }\rightarrow V\left( u\right) \in T_u\left( {\cal E_N}%
\right) .
$$
Vectors $\left( \partial _\alpha \right) _u,(\alpha =0,1,2,...,m+n-1),$ form
a local linear basis in $T_u\left( {\cal E_N}\right) .$ We shall also use
decompositions on locally adapted basis (2.4), $\left( \delta _\alpha
\right) ,$ and denote by {\bf X$\left( {\cal E_N}\right) $} the set of $%
C^\infty $-vector fields on ${\cal E_N.}$

Now, let introduce the bundle of linear adapted frames $GL\left( {\cal E_N}%
\right) $ on ${\cal E_N.}$ As a linear adapted frame $e=[e_{\underline{%
\alpha }}],(\underline{\alpha }=0,1,...,m+n-1),$ in point $u\in {\cal E_N}$
we mean a linear independent system of vectors $e_{\underline{\alpha }}\in
T_u\left( {\cal E_N}\right) $ obtained as a linear distinguished transform
of local adapted basis (2.4), i.e.$e_{\underline{\alpha }}=e_{\underline{%
\alpha }}^\alpha \delta _\alpha ,$ where $e_{\underline{\alpha }}^\alpha \in
GL^d\left( {\cal R}\right) =GL\left( n,{\cal R\ }\right) \oplus GL\left( m,%
{\cal R}\right) .$ Then $GL\left( {\cal E_N}\right) $ is defined as the set
of all linear adapted frames in all points $u\in {\cal E_N:}$

$GL\left( {\cal E_N}\right) =\{r=(u,e),u\in {\cal E_N}$ and $e$ is a linear
adapted frame in the point $u\}.$

Local coordinate carts on $GL\left( {\cal E_N}\right) $ are defined as $%
\left( \widetilde{{\cal U\ }}_\alpha ,\widetilde{\varphi }_\alpha \right) ,$
where $\widetilde{{\cal U\ }}_\alpha =\{r=(u,e)\in GL\left( {\cal E_N}%
\right) ,u\in {\cal U_\alpha \ \subset E_{N,}}$ and $e$ is a linear adapted
frame in the point $u\}$ , $\widetilde{\varphi _\alpha }\left( r\right)
=\left( \varphi _\alpha \left( u\right) =(u^\alpha ),e_{\underline{\alpha }%
}^\alpha \right) $ and $e_{\underline{\alpha }}=e_{\underline{\alpha }%
}^\alpha \delta _\alpha \mid _u.$ So $GL\left( {\cal E_N}\right) $ has the
structure of $C^\infty $-manifold of dimension $n+m+n^2+m^2.$ Elements $a\in
GL^d\left( {\cal R}\right) $ act on $GL\left( {\cal E_N}\right) $ according
the formula $T_a(u,e)=(u,ea),$ where $(ea)_{\underline{\alpha }}=a_{
\underline{\alpha }}^{\underline{\beta }}e_{\underline{\beta }}.$ The
surjective projection $\pi :GL\left( {\cal E_N}\right) \rightarrow {\cal %
E_N\ }$is defined in a usual manner by the equality $\pi (u,e)=u.$

Every vector field $L\in {\bf X\left( {\cal E_N}\right) }$ induces a vector
field $\widetilde{L}$ on $GL\left( {\cal E_N}\right) .$ Really, for $f\in
{\bf X\left( {\cal E_N}\right) }$ we can consider
\begin{equation}
\left( \widetilde{L}f\right) (r)=\frac d{dt}f\left( (\exp tL)u,(\exp
tL)_{*}e\right) \mid _{t=0},\eqnum{2.18}
\end{equation}
where $r=(u,e)$ and
$$
(\exp tL)_{*}e=[(\exp tL)_{*}e_1,(\exp tL)_{*}e_2,...,(\exp tL)_{*}e_{m+n}],
$$
is the differential (an isomorphism $T_u\left( {\cal E_N\ }\right)
\rightarrow T_{(\exp tL)_u}{\cal E_N}$ for every $u\in {\cal E_N)}$ of $\exp
tL$ and the local diffeomorphism $u\rightarrow v(t,u)$ is defined by
differential equations%
\begin{equation}
\frac{dv^\alpha }{dt}(t,u)=a^\alpha (v(t,u)),\eqnum{2.19}
\end{equation}
$$
\left( L=a^\alpha (u)\delta _\alpha \right) ,v(0,u)=u.
$$

Let $L\in {\bf X\left( {\cal E_N}\right) }$ and introduce functions $%
f_L^\alpha (r)\in F\left( GL({\cal E_N)}\right) $ for every $\alpha
=0,1,2,...,m+n-1$ by the equalities%
\begin{equation}
f_L^\alpha (r)=\left( e^{-1}\right) _{\underline{\alpha }}^\alpha a^{
\underline{\alpha }}(u)\eqnum{2.20}
\end{equation}
written in locally adapted coordinates $r=\left( u^\alpha ,e=(e_{\underline{%
\alpha }}^\alpha )\right) $ on the manifold $GL\left( {\cal E_N}\right) ,$
where $L=a^\alpha (u)\delta _\alpha $ and $e^{-1}$ is the matrix inverse to $%
e.$ Because the equality (2.20)does not depend on local coordinates, we have
defined a global function on $GL\left( {\cal E_N}\right) .$ It's obvious
that for $L_{(1)},L_{(2)}\in {\bf X\left( {\cal E_N}\right) }$ we have
$$
\left( \widetilde{L}_{(1)}f_{L_{(2)}}^\alpha \right) \left( r\right)
=f_{[L_{(1)}L_{(2)}]}^\alpha \left( r\right) ,
$$
where $\widetilde{L}_{(1)}$ and $\widetilde{L}_{(2)}$ are constructed
similarly to operator (2.18), and%
$$
[L_{(1)},L_{(2)}]=L_{(1)}L_{(2)}-L_{(2)}L_{(1)}.
$$
A distinguished connection $\Gamma _{\beta \gamma }^\alpha $ defines the
covariant derivation of d-tensors in ${\cal E_N}$ in a usual manner. For
example, we can introduce a d-covariant derivation $DB$ of a d-tensor field $%
B\left( u\right) =B_{\beta _1\beta _2...\beta _q}^{\alpha _1\alpha
_2...\alpha _p}\left( u\right) $ in the form%
$$
D_\gamma B_{\beta _1\beta _2...\beta _q}^{\alpha _1\alpha _2...\alpha
_p}(u)=B_{\beta _1\beta _2...\beta _q;\gamma }^{\alpha _1\alpha _2...\alpha
_p}(u)=\delta _\gamma B_{\beta _1\beta _2...\beta _q}^{\alpha _1\alpha
_2...\alpha _p}(u)+
$$
\begin{equation}
\sum_{\epsilon =1}^p\Gamma _{\gamma \delta }^{\alpha _\epsilon }\left(
u\right) B_{\beta _1\beta _2...\beta _q}^{\alpha _1\alpha _2...\delta
...\alpha _p}(u)-\sum_{\tau =1}^q\Gamma _{\gamma \beta _\tau }^\delta \left(
u\right) B_{\beta _1\beta _2...\delta ...\beta _q}^{\alpha _1\alpha
_2...\alpha _p}(u),\eqnum{2.21}
\end{equation}
or the covariant derivative $D_YB$ in the direction $Y=Y^\alpha \delta
_\alpha \in {\bf X\left( {\cal E_N\ }\right) }$ ,%
\begin{equation}
(D_YB)_{\beta _1\beta _2...\beta _q}^{\alpha _1\alpha _2...\alpha
_p}(u)=Y^\delta B_{\beta _1\beta _2...\beta _q;\delta }^{\alpha _1\alpha
_2...\alpha _p}(u)\eqnum{2.22}
\end{equation}
and the parallel transport along a (piecewise) smooth curve $c:{\cal R\
\supset \,}I=(t_{1,}t_2)\ni t\rightarrow c\left( t\right) $ (considering $%
B\left( t\right) =B\left( c(t)\right) $ )%
$$
\frac d{dt}B_{\beta _1\beta _2...\beta _q}^{\alpha _1\alpha _2...\alpha
_p}(t)+\sum_{\epsilon =1}^p\Gamma _{\gamma \delta }^{\alpha _\epsilon
}\left( c(t)\right) B_{\beta _1\beta _2...\beta _q}^{\alpha _1\alpha
_2...\delta ...\alpha _p}(c(t))\frac{dc^\gamma }{dt}-
$$
\begin{equation}
\sum_{\tau =1}^q\Gamma _{\gamma \beta _\tau }^\delta \left( c(t)\right)
B_{\beta _1\beta _2...\delta ...\beta _q}^{\alpha _1\alpha _2...\alpha
_p}(c(t))\frac{dc^\gamma }{dt}=0.\eqnum{2.23}
\end{equation}
For every $r\in GL\left( {\cal E_L}\right) $ we can define the horizontal
subspace
$$
H_r=\{U=a^\alpha \delta _\alpha \mid _u-\Gamma _{\beta \gamma }^\alpha
\left( u\right) e_{\underline{\gamma }}^\gamma a^\beta \frac \partial
{\partial e_{\underline{\gamma }}^\alpha },a^\alpha \in {\cal R}^{m+n}\}
$$
of $T_u\left( GL({\cal E_N)}\right) .$ Vector $U\in H_r\,$ is called
horizontal. Let $\xi \in T_u\left( {\cal E_N\ }\right) ,$ then $\widetilde{%
\xi }\in T_r\left( GL({\cal E_N)}\right) \,$is a horizontal lift of $\xi $
if the vector $\widetilde{\xi }$ is horizontal, i.e. $\pi \left( r\right) =u$
and $(d\pi )_r\widetilde{\xi }=\xi .$ If $r$ is given as to satisfy $\pi
\left( r\right) =u$ the $\widetilde{\xi \text{ }}$ is uniquely defined. So,
for given $U\in {\bf X\left( {\cal E_N}\right) \ }$there is a unique $
\widetilde{U}\in {\bf X\ }\left( GL\left( {\cal E_N}\right) \right) ,$ where
$\widetilde{U}_r$ is the horizontal lift $U_{\pi \left( r\right) }$ for
every $r\in GL\left( {\cal E_N}\right) .\widetilde{U}$ is called the
horizontal lift of vector field $U.$ In local coordinates $\widetilde{U}%
=U^\alpha \left( u\right) \delta _\alpha -\Gamma _{\alpha \beta }^\delta
\left( u\right) U^\alpha \left( u\right) e_{\underline{\epsilon }}^\beta
\frac \partial {\partial e_{\underline{\epsilon }}^\delta }$ if $U=U^\alpha
\left( u\right) \delta _\alpha .$

In a sim\-i\-lar man\-ner we can de\-fine the hor\-i\-zon\-tal lift
$$\widetilde{c}\left(
t\right) =\left( c(t),e(t)\right) =[e_0(t),e_1(t),...,e_{q-1}(t)]\in
GL\left( {\cal E_N}\right) $$ of a curve $c\left( t\right) \in {\cal E_N}$
with the property that $\pi \left( \widetilde{c}(t)\right) =c(t)\,$ for $%
t\in I\,$ and $\frac{d\widetilde{c}}{dt}(t)$ is horizontal. For every $
\underline{\alpha }=0,1,...q-1$ there is a unique vector field $\widetilde{L}%
\in {\bf X}\left( GL({\cal E_N)}\right) ,$ for which $\left( \widetilde{L}_{
\underline{\alpha }}\right) _r$ is the horizontal lift of vector $e_{
\underline{\alpha }}\in T_u\left( {\cal E_N}\right) $ for every $r=\left(
u,e=\left[ e_0,e_1,...,e_{q-1}\right] \right) .$ In coordinates $\left(
u^\alpha ,e_{\underline{\beta }}^\beta \right) $ we can express%
\begin{equation}
\widetilde{L}_{\underline{\alpha }}=e_{\underline{\alpha }}^\alpha \delta
_\alpha -\Gamma _{\beta \gamma }^\alpha e_{\underline{\alpha }}^\beta e_{
\underline{\beta }}^\gamma \frac \partial {\partial e_{\underline{\beta }%
}^\alpha }.\eqnum{2.24}
\end{equation}
Vector fields $\widetilde{L}_{\underline{\alpha }}$ form the system of
canonical horizontal vector fields.

Let $B\left( u\right) =B_{\beta _1\beta _2...\beta _s}^{\alpha _1\alpha
_2...\alpha _p}\left( u\right) $ be a (p,s)-tensor field and define a system
of smooth functions
$$
F_B\left( r\right) =\{F_{B\underline{\beta _1}\underline{\beta _2}...
\underline{\beta _s}}^{\underline{\alpha _1}\underline{\alpha _2}...
\underline{\alpha _p}}\left( r\right) =B_{\delta _1\delta _2...\delta
_s}^{\gamma _1\gamma _2...\gamma _p}\left( u\right) e_{\gamma _1}^{
\underline{\alpha _1}}e_{\gamma _2}^{\underline{\alpha _2}}...e_{\gamma _p}^{
\underline{\alpha _p}}e_{\underline{\beta _1}}^{\delta _1}e_{\underline{%
\beta _2}}^{\delta _2}...e_{\underline{\beta _s}}^{\delta _s}\}
$$
(the scalarization of the d-tensor field $B\left( r\right) $ with the
respect to the locally adapted basis $e)$ on $GL\left( {\cal E_N}\right) ,$
where we consider that%
$$
B\left( u\right) =F_{B\underline{\beta _1}\underline{\beta _2}...\underline{%
\beta _s}}^{\underline{\alpha _1}\underline{\alpha _2}...\underline{\alpha _p%
}}\left( u\right) e_{\underline{\alpha _1}}\otimes e_{\underline{\alpha _2}%
}\otimes ...e_{\underline{\alpha _p}}\otimes e_{*}^{\underline{\beta _1}%
}\otimes e_{*}^{\underline{\beta _2}}...\otimes e_{*}^{\underline{\beta _s}%
},
$$
the matrix $e_{\underline{\beta }}^\delta $ is inverse to the matrix $%
e_\gamma ^{\underline{\alpha }},$ basis $e_{*}$ is dual to $e$ and $r=(u,e).$
It is easy to verify that
\begin{equation}
\widetilde{L}_{\underline{\alpha }}(F_{B\underline{\beta _1}\underline{\beta
_2}...\underline{\beta _s}}^{\underline{\alpha _1}\underline{\alpha _2}...
\underline{\alpha _p}})\left( r\right) =(F_{\nabla B})_{\underline{\beta _1}
\underline{\beta _2}...\underline{\beta _s};\underline{\alpha }}^{\underline{%
\alpha _1}\underline{\alpha _2}...\underline{\alpha _p}}\left( r\right)
\eqnum{2.25}
\end{equation}
where the covariant derivation $\nabla _{\underline{\alpha }}A^{\underline{%
\beta }}=A_{;\underline{\alpha }}^{\underline{\beta }}$ is taken by using
connection%
$$
\Gamma _{\underline{\beta }\underline{\gamma }}^{\underline{\alpha }%
}=e_\alpha ^{\underline{\alpha }}e_{\underline{\beta }}^\beta e_{\underline{%
\gamma }}^\gamma \Gamma _{\beta \gamma }^\alpha +e_{\underline{\gamma }%
}^\gamma e_\sigma ^{\underline{\alpha }}\delta _\gamma e_{\underline{\beta }%
}^\sigma
$$
in $GL\left( {\cal E_N}\right) $ (induced from ${\cal E_N).}$

In our further considerations we shall also use the bundle of
orthonormalized adapted frames on ${\cal E_N,\ }$defined as a subbundle of $%
GL\left( {\cal E_N}\right) $ satisfying conditions:

$O\left( {\cal E_N}\right) =\{r=(u,e)\in GL\left( {\cal E_N}\right) ,e$ is a
orthonormalized basis in $T_u\left( {\cal E_N}\right) \}.$

So $r=(u^\alpha ,e_{\underline{\alpha }}^\alpha \in O\left( {\cal E_N}%
\right) )$ if and only if%
\begin{equation}
G_{\alpha \beta }e_{\underline{\alpha }}^\alpha e_{\underline{\beta }}^\beta
=\delta _{\underline{\alpha }\underline{\beta }}\eqnum{2.26}
\end{equation}
or, equivalently,%
$$
\sum_{\underline{\alpha }=0}^{q-1}e_{\underline{\alpha }}^\alpha e_{
\underline{\alpha }}^\beta =G^{\alpha \beta },
$$
where the matrix $G^{\alpha \beta }$ is inverse to the matrix $G_{\alpha
\beta }$ from (2.10).

\section{Stochastic Differential Equations in Vector Bundle Spaces}

In this Section we assume that the reader is familiar with the concepts and
basic results on stochastic calculus, Brownian motion and diffusion
processes (an excellent presentation can be found in [7-9, 18-20], see also a
brief introduction into the mentioned subjects in the Appendix of this
paper). The purpose of the Section is to extend the theory of stochastic
differential equations on Riemannian spaces [7-9] to the case of spaces with
general anisotropy, defined in the previous Section as v-bundles.

Let $A_{\widehat{0}},A_{\widehat{1}},...,A_{\widehat{r}}\in {\bf X\left(
{\cal E_N}\right) }$ and consider stochastic differential equations%
\begin{equation}
dU\left( t\right) =A_{\widehat{\alpha }}\circ dB^{\widehat{\alpha }}\left(
t\right) +A_{\widehat{0}}\left( U(t)\right) dt,\eqnum{3.1}
\end{equation}
where $\widehat{\alpha }=1,2,...,r$ and $\circ \,$ is the symmetric
Q-product (see Appendix A1). We shall use the point compactification of
space ${\cal E_N}$ and write $\widehat{{\cal E\ }}_N={\cal E\ }$ or $
\widehat{{\cal E\ }}_N={\cal E\ }$ $\cup \{\Delta \}$ in dependence of that
if ${\cal E_N\ }$ is compact or noncompact. By $\widehat{W}\left( {\cal E_N}%
\right) $ we denote the space of paths in ${\cal E_N\ ,\ }$ defined as

$\widehat{W}\left( {\cal E_N}\right) =\{w:w$ is a smooth map [0,$\infty
)\rightarrow \widehat{\cal E\ }_N$ with the property that $w\left( 0\right)
\in {\cal E_N}$ and $w(t)=\Delta ,w(t^{\prime })=\Delta $ for all $t^{\prime
}\ge t\}$

and by ${\cal B}$ $\left( \widehat{W}\left( {\cal E_N}\right) \right) $ the $%
\sigma $-field generated by Borel cylindrical sets.

The explosion moment $e(w)$ is defined as
$$
e\left( w\right) =\inf \{t,w(t)=\Delta \}
$$

{\bf Definition 3.1.}{\it The solution }$U=U\left( t\right) $ {\it of
equation (3.1) in v-bundle space }${\cal E_N\ }$ {\it is defined as such a }$%
{( {\cal F}_t)}$\text{{\it -compatible }}$\widehat{W}\left( {\cal E_N}\right) $-%
{\it valued random element (i.e. as a smooth process in }${\cal E_N\ }$ {\it %
with the trap $\Delta ),$} {\it given on the probability space with
filtration } $({\cal F}_t)$ {\it and } $r$-{\it dimensional } ${\cal F}_t)$-%
{\it Brownian motion} $B=B(t),$ with $B(0)=0,$ for which

$$
f\left( U\left( t\right) \right) -f\left( U\left( 0\right) \right) =
$$
\begin{equation}
\int_0^tA_{\widehat{\alpha }}\left( t\right) \left( U\left( s\right) \right)
\delta B^{\widehat{\alpha }}\left( s\right) +\int_0^t\left( A\circ f\right)
\left( U\left( s\right) \right) ds \eqnum{3.2}
\end{equation}
{\it for every }$f\in F_0\left( {\cal E_N}\right) $ {\it (we consider }$%
f\left( \Delta \right) =0),$ {\it \ where the first term is understood as a
Fisk-Stratonovich integral.}

In (3.2) we use $\delta B^{\widehat{\alpha }}\left( s\right) $ instead of $%
dB^{\widehat{\alpha }}\left( s\right) $ because on ${\cal E_N}$ the Brownian
motion must be adapted to the N-connection structure.

In a manner similar to that for stochastic equations on Riemannian spaces
[7] we can construct the unique strong solution to the equations (3.1). To
do this we have to use the space of paths in ${\cal R}^r$ starting in point
0, denoted as $W_0^r,$ the Wiener measure $P^W$ on $W_0^r,\sigma $-field $%
{\cal B}_t\ (W_0^r)$-generated by Borel cylindrical sets up to moment $t$ and
the similarly defined $\sigma $-field.

{\bf Theorem 3.1.}{\it \ There is a function }$F:{\cal E_N\ }\times
W_0^r\rightarrow \widehat{W}\left( {\cal E_N\ }\right) $ {\it being} $%
\bigcap_\mu {\cal B} \left( {\cal E}_N\right) \times {\cal B}_t\ $  $\left(
W_0^r\right) ^{\mu \times P^W}/${\it ${\cal B}_t\ $} $\left( \widehat{W}%
\left( {\cal F}_t \right) \right) $-{\it measurable (index $\mu $} {\it runs
all probabilities in $\left( {\cal E_N,B(E_N)\ }\right) $} ) {\it for every }%
$t\geq 0$ {\it and having properties:}

1){\it For every }$U(t)$ {\it and Brownian motion }$B=B\left( t\right) $
{\it the equality }$U=F(U\left( 0\right) ,B)$ {\it a.s. is satisfied.}

2){\it For every r-dimensional } $({\cal F}_t )$-{\it Brownian motion }$%
B=B\left( t\right) $ {\it with } $B=B\left( 0\right) ,$ {\it defined on the
probability space with filtration } ${\cal F}_t,\ $\text{{\it and }} ${\cal %
E_N}$-{\it valued }${\cal F_0}$-{\it measurable random element $\xi ,$} {\it %
the function }$U=F(\xi ,B)$ {\it is the solution of the differential
equation (3.1) with }$U\left( 0\right) =\xi ,$ {\it a.s.}

{\bf Sketch of the proof.} Let take a compact coordinate vicinity ${\cal V }$%
with respect to a locally adapted basis $\delta _\alpha $ and express $A_{
\underline{\alpha }}=\sigma _{\underline{\alpha }}^\alpha \left( u\right)
\delta _\alpha ,$ where functions $\sigma _{\underline{\alpha }}^\alpha
\left( u\right) $ are considered as bounded smooth functions in ${\cal %
R}^{m+n}$ , and consider on ${\cal V}$ the stochastic differential equation%
\begin{equation}
dU_t^\alpha =\sigma _{\widehat{\alpha }}^\alpha \left( U_t^\alpha \right)
\circ \delta B^{\widehat{\alpha }}\left( t\right) +\sigma _0^\alpha \left(
U_t\right) dt,\eqnum{3.3}
\end{equation}
$$
U_0^\alpha =u^\alpha ,(\alpha =0,1,...q-1).
$$
Equations (3.3) are equivalent to%
$$
dU_t^\alpha =\sigma _{\widehat{\alpha }}^\alpha \left( U_t^\alpha \right)
dB^{\widehat{\alpha }}\left( t\right) +\overline{\sigma }_0^\alpha \left(
U_t\right) dt,
$$
$$
U_0^\alpha =u^\alpha ,
$$
where $\overline{\sigma }_0^\alpha \left( u\right) =\sigma _0^\alpha \left(
u\right) +\frac 12\sum_{\widehat{\alpha }=1}^r\left( \frac{\delta \sigma _{
\widehat{\alpha }}^\alpha \left( u\right) }{\delta u^\beta }\right) \sigma
_\beta ^{\widehat{\alpha }}\left( u\right) .$ It's known [7] that (3.3) has
a unique strong solution $F:{\cal R}^{n+m}\ \times W_0\rightarrow \widehat{W}%
^{n+m}$ or $F(u,w)=\left( U(t,u,w)\right) .$ Taking $\tau _{{\cal V}%
}(w)=\inf \{t:U(t,u,w)\in {\cal V\ \}}$ we define
\begin{equation}
U_{{\cal V\ }}(t,u,w)=U(t\bigwedge \tau _{{\cal V\ }}\left( w\right)
,u,w)\eqnum{3.4}
\end{equation}

In a point $u\in {\cal V\cap \widetilde{V},}$ where ${\cal \widetilde{V}\ }$
is covered by local coordinates $u^{\widetilde{\alpha }},$ we have to
consider transformations
$$
\sigma _{\underline{\alpha }}^{\widetilde{\alpha }}\left( \widetilde{u}%
\left( u\right) \right) =\sigma _{\underline{\alpha }}^\alpha (u)\frac{%
\partial u^{\widetilde{\alpha }}}{\partial u^\alpha },
$$
where coordinate transforms $u^{\widetilde{\alpha }}\left( u^\alpha \right) $
satisfy the properties (2.1). The global solution of (3.3) can be
constructed by gluing together functions (3.4) defined on corresponding
coordinate regions covering ${\cal E_N\ }$. $\diamondsuit $ (end of the
proof)

Let $P_u$ be a probability law on $\widehat{W}\left( {\cal E_N}\right) $ of
a solution $U=U\left( t\right) $ of equation (3.1) with initial conditions $%
U_{(0)}=u.$ Taking into account the uniqueness of the mentioned solution we
can prove that $U=U\left( t\right) $ is a A-diffusion and satisfy the Markov
property [7] (see also Appendix A3). Really, because for every $f\in
F_0\left( {\cal E_N}\right) $

$$
df\left( U\left( t\right) \right) =\left( A_{\widehat{\alpha }}f\right)
\left( U\left( t\right) \right) \circ dw^{\widehat{\alpha }}+\left(
A_0f\right) \left( U\left( t\right) \right) dt=
$$

$$
\left( A_{\widehat{\alpha }}f\right) \left( U\left( t\right) \right) dw^{
\widehat{\alpha }}+\left( A_0f\right) \left( U\left( t\right) \right)
d+\frac 12d\left( A_{\widehat{\alpha }}f\right) \left( U\left( t\right)
\right) \cdot dw^{\widehat{\alpha }}\left( t\right)
$$
and%
$$
d\left( A_{\widehat{\beta }}f\right) \left( U\left( t\right) \right) =A_{
\widehat{\alpha }}\left( A_{\widehat{\beta }}f\right) \left( U\left(
t\right) \right) \circ dw^{\widehat{\alpha }}\left( t\right) +\left( A_{
\widehat{0}}A_{\widehat{\beta }}f\right) \left( U\left( t\right) \right) dt,
$$
we have%
$$
d\left( A_{\widehat{\alpha }}f\right) \left( U\left( t\right) \right) \cdot
dw^{\widehat{\alpha }}\left( t\right) =\sum\limits_{\widehat{\alpha }%
=1}^rA_{ \widehat{\alpha }}\left( A_{\widehat{\alpha }}f\right) \left(
U\left( t\right) \right) dt.
$$
Consequently, it follows that%
$$
df\left( U\left( t\right) \right) =\left( A_{\widehat{\alpha }}f\right)
\left( U\left( t\right) \right) dw^{\widehat{\alpha }}\left( t\right)
+\left( Af\right) \left( U\left( t\right) \right) dt,
$$
i.e. the operator $\left( Af\right) ,$ defined by the equality
\begin{equation}
Af=\frac 12\sum\limits_{\widehat{\alpha }=1}^rA_{\widehat{\alpha }}\left(
A_{ \widehat{\alpha }}f\right) +A_0f\eqnum{3.5}
\end{equation}
generates a diffusion process $\{P_u\},u\in {\cal E_N.}$

The above presented results are summarized in this form:

{\bf Theorem 3.2. }{\it A second order differential operator }$Af\,$ {\it %
generates a A-diffusion on $\widehat{W}$}$({\cal E_N)}$ {\it of a solution }$%
U=U\left( t\right) \,\,$ {\it of the equation (3.5) with initial condition }$%
U\left( 0\right) =u.\,$

Using similar considerations as in flat spaces [7] on carts covering ${\cal %
E_{N.},\ }$ we can prove the uniqueness of A-diffusion$\{P_u\},u\in {\cal %
E_N\ }$ on {\it $\widehat{W}$}$({\cal E_N).}$

\section{Heat  Equations in Bundle Spaces and Flows of Diffeomorfisms}

Let v-bundle ${\cal E_N\ }$ be a com\-pact man\-i\-fold of class $C^\infty .\,$We
con\-sider op\-er\-a\-tors
$$A_0,A_1,...,A_r\in {\bf X\left( {\cal E_N}\right) }$$ and
suppose that the property $$E[Sup_{t\in [0,1]}Sup_{u\in {\cal U\ }}|D^{
\underline{\alpha }}\{f\left( U\left( t,u,w\right) \right) \}|]<\infty $$ is
satisfied for all $f\in F_0\left( {\cal E_N}\right) $ and every multiindex
$\underline{\alpha}$ in the coordinate vicinity ${\cal U\ }$ with $\overline{%
{\cal U}}$ being compact for every $T>0.$ The heat equation in $F_0\left(
{\cal E_N}\right) {\cal _N}$ is written as
\begin{equation}
\frac{\partial \nu }{\partial t}(t,u)=A\nu (t,u),\eqnum{4.1}
\end{equation}
$$
\lim \limits_{t\downarrow 0,\overline{u}\rightarrow u}\nu (t,\overline{u}%
)=f\left( u\right) ,
$$
where operator $A$ acting on $F_{}\left( F_0\left( {\cal E_N}\right) \right)
$ $\,$ is defined in (3.5).

We denote by $C^{1,2}\left( [0,\infty )\times F_0\left( {\cal E_N}\right)
\right) $ the set of all functions $f(t,u)$ on $[0,\infty )\times {\cal E_N}$
being smoothly differentiable on $t$ and twice differentiable on $u.$

The existence and properties of solutions of equations (4.1) are stated
according the theorem:

{\bf Theorem 4.1. }{\it The function
\begin{equation}
\zeta (t,u)=E[f(U(t,u,w)]\in C^\infty \left( [0.\infty )\times {\cal E_N}%
\right) \eqnum{4.2}
\end{equation}
}$f\in F_0\left( {\cal E_N}\right) $ {\it satisfies heat equation (4.1).
Inversely, if a bounded function} $\nu (t,u)\in C^{1,2}\left( [0,\infty
)\times {\cal E_N}\right) $ {\it solves equation (4.1) and satisfies the
condition}%
\begin{equation}
\lim \limits_{k\uparrow \infty }E[\nu (t-\sigma _k,U\left( \sigma
_k,u,w\right) ):\sigma _k\leq t]=0 \eqnum{4.3}
\end{equation}
{\it for every} $t>0$ {\it and} $u\in {\cal E_N,\ }$ {\it where} $\sigma
_k=\inf \{t,U(t,u,w)\in D_k\}$ {\it and} $D_k$ {\it is an increasing
sequence with respect to closed sets in} ${\cal E_N,}$ $\bigcup\limits_kD_k=$
${\cal E_N.}$

{\rm Sketch of the proof. }The function $\zeta (t,u)$ is a function on $%
{\cal E_N,\ }$ because $u\rightarrow f\left( U\left( t,u,w\right) \right)
\in C^\infty $ and in this case the derivation under mathematical
expectation symbol is possible. According to (3.2) we have%
$$
f\left( U\left( t,u,w\right) \right) -f\left( u\right)
=martingale+\int_0^t\left( At\right) \left( U\left( s,u,w\right) \right) ds
$$
for every $u\in {\cal E_N,\ }$ i.e.%
\begin{equation}
\zeta (t,u)=f\left( u\right) +\int_0^tE[\left( Af\right) \left( U\left(
s,u,w\right) \right) ds. \eqnum{4.4}
\end{equation}
Because $A^nf\in F_0\left( {\cal E_N}\right) ,\left( n=1,2,...\right) ,$ we
can write%
$$
\zeta (t,u)=f\left( u\right) +t\left( Af\right) \left( u\right)
+\int_0^tdt_1\int_0^{t_1}E\left[ \left( A^2f\right) \left( U\left(
t_2,u,w\right) \right) \right] dt_2=
$$
$$
f\left( u\right) +t\left( Af\right) \left( u\right) +\frac{t^2}2\left(
A^2f\right) \left( u\right)
+\int_0^tdt_1\int_0^{t_1}dt_2\int_0^{t_2}E[\left( A^3f\right) \left( U\left(
t_3,u,w\right) \right) ]dt_3=
$$
$$
f\left( u\right) +t\left( Af\right) \left( u\right) +\frac{t^2}2\left(
A^2f\right) \left( u\right)
+...+\int_0^tdt_1\int_0^{t_1}dt_2...\int_0^{t_{n-1}}E[\left( A^nf\right)
\left( U\left( t_n,u,w\right) \right) dt_n
$$
from which it is clear that

$$
\zeta \left( t,u\right) \in C^\infty \left( [0,\infty )\times {\cal E_N}%
\right) .
$$
In Chapter V, Section 3 of the monograph [7] it is proved the equality
\begin{equation}
\left( A\zeta _t\right) \left( u\right) =E[\left( Af)U\left( t,u,w\right)
\right) ] \eqnum{4.5}
\end{equation}
for every $t\geq 0,$ where $\zeta _t\left( u\right) =\zeta \left( t,u\right)
.$

From (4.4) and (4.5) one follows that
$$
\zeta \left( t,u\right) =f\left( u\right) +\int_0^t\left( A\zeta \right)
\left( s,u\right) ds
$$
and%
$$
\frac{\partial \zeta }{\partial t}\left( t,u\right) =A\zeta \left(
t,u\right) ,
$$
i.e. $\zeta =\zeta \left( t,u\right) $ satisfies the heat equation (4,1).

Inversely, let $\nu \left( t,u\right) $ $\in C^{1,2}\left( [0,\infty )\times
{\cal E_N}\right) $ be a bounded solution of the equation (4.1). Taking into
account that $P\left( e[U\left( \cdot ,u,w\right) \right) ]=\infty )=1$ for
every $u\in {\cal E}_N$ and using the Ito formula (see (A6) in the Appendix)
we obtain that for every $t_0>0$ and $0\leq t\leq t_0\,$ .%
$$
E[\nu \left( t_0-t\bigwedge \sigma _n,U\left( t\bigwedge \sigma
_n,u,w\right) \right) ]-\nu \left( t_0,u\right) =
$$
\begin{equation}
E[\int_0^{t\bigwedge \sigma _n}\{\left( A\nu \right) \left( t_0-s,U\left(
s,u,w\right) \right) -\frac{\partial \nu }{\partial t}\left( t_0-s,U\left(
s,u,w\right) \right) \}ds]. \eqnum{4.6}
\end{equation}
Supposing that conditions (4.3) are satisfied and considering $n\uparrow
\infty $ we obtain
$$
E=\{\nu \left( t_0-t,U\left( t,u,w\right) \right) ;e[U\left( \cdot
,u,w\right) ]>t\}=\nu \left( t_0,u\right) .
$$
For $t\uparrow t_0$ we have $E[f\left( U\left( t_0,u,w\right) \right)
]=\zeta \left( t_0,u\right) ,$ i.e. $\nu \left( t,u\right) =\zeta \left(
t,u\right) .\diamondsuit $

{\bf Remarks; 1.}{\it \ The conditions (4.3) are necessary in order in order
to select a unique solution of (4.1).}

2.{\it \ Defining
$$
\zeta \left( t,u\right) =E[\exp \{\int_0^tC\left( U\left( s,u,w\right)
\right) ds\}f\left( U\left( t,u,w\right) \right) ]
$$
instead of (4.2) we generate the solution of the generalized heat equation
in }${\cal E_N:}$
$$
\frac{\partial \nu }{\partial t}\left( t,u\right) =\left( A\nu \right)
\left( t,u\right) +C\left( u\right) \nu \left( t,u\right) ,
$$
$$
\lim \limits_{t\downarrow 0,\overline{u}\rightarrow u}\nu \left( t,\overline{%
u}\right) =f\left( u\right) .
$$

For given vector fields $A_{\left( \alpha \right) }\in {\bf X}{\cal \left(
E_N\right) },\left( \alpha \right) =0,1,...,r$ in Section III we have
constructed the map%
$$
U=\left( U\left( t,u,w\right) \right) :{\cal E_N\ \times }W_0^r\ni
(u,w)\rightarrow U\left( \cdot ,u,w\right) \in \widehat{W}{\cal \left(
E_N\right) ,}
$$
which can be constructed as a map of type
$$
[0,\infty )\times {\cal E_N\ \times }W_0^r\ni (u,w)\rightarrow U\left(
t,u,w\right) \in \widehat{{\cal E_N\ }}.
$$
Let us show that map $u\in {\cal E_N\ }\rightarrow U\left( t,u,w\right) \in
\widehat{{\cal E_N\ }}$ is a local diffeomorphism of the manifold ${\cal E_N}$
for every fixed $t\geq 0$ and almost every $w$ that $\ \in {\cal E_N\ }$ .

We first consider the case when ${\cal E_N\ }$ $\cong {\cal R}^{n+m},\sigma
\left( u\right) =\left( \sigma _\beta ^\alpha \left( u\right) \right) \in
{\cal R}^{n+m}\otimes{\cal R}^{n+m}$ and $b\left( u\right) =\left(
b^\alpha \left( u\right) \right) \in {\cal R}^{n+m}$ are given smooth
functions (i.e. $C^\infty $-functions) on ${\cal R}^{n+m}$ , $\parallel
\sigma \left( u\right) \parallel +\parallel b\left( u\right) \parallel \leq
K\left( 1+|u|\right) $ for a constant $K>0$ and all derivations of $\sigma
^\alpha $ and $b^\alpha \,$ are bounded. It is known [7] that there is a
unique solution $U=U\left( t,u,w\right) $ , with the property that $E[\left(
U\left( t\right) \right) ^p]<\infty $ for all $p>1,$ of the equation%
\begin{equation}
dU_t^\alpha =\sigma _{\widehat{\alpha }}^\alpha \left( U_t\right) dw^{
\widehat{\alpha }}\left( t\right) +b^\alpha \left( U_t\right) dt, \eqnum{4.7}
\end{equation}
$$
U_0=u,(\alpha =1,2,...,m+n-1),
$$
defined on the space $\left( W_0^r,P^W\right) $ with the flow $\left( {\cal %
F}_t^0\right) .$

In order to show that the map $u\rightarrow U\left( t,u,w\right) $ is a
diffeomorphism of ${\cal R}^{n+m}$ it is more convenient to use the
Fisk-Stratonovich differential (see, for example, the Appendix A1) and to
write the equation (4.7) equivalently as
\begin{equation}
dU_t^\alpha =\sigma _{\widehat{\alpha }}^\alpha \left( U_t\right) \circ
\delta w^{\widehat{\alpha }}\left( t\right) +\overline{b}^\alpha \left(
U_t\right) dt,\eqnum{4.8}
\end{equation}
$$
U_0=u,
$$
by considering that
\begin{equation}
\overline{b}^\alpha \left( u\right) =b^\alpha \left( u\right) +\frac
12\sum_{ \widehat{\alpha }=1}^r\left( \delta _\beta \sigma _{\widehat{\alpha
}}^\alpha \right) \sigma _{\widehat{\alpha }}^\beta \left( u\right) .\eqnum{4.9}
\end{equation}
We emphasize that for solutions of equations of type (4.8) one holds the
usual derivation rules as in mathematical analysis.

Let introduce matrices $\sigma _{\widehat{\alpha }}^{\prime }=\left( \sigma
^{\prime }\left( u\right) _{\widehat{\alpha }\beta }^\alpha =\frac \delta
{\delta u^\beta }\sigma _{\widehat{\alpha }}^\alpha \left( u\right) \right)
,b^{\prime }\left( u\right) =\left( b^{\prime }\left( u\right) _\beta
^\alpha =\frac{\delta b^\alpha }{\delta u^\beta }\right) ,I=\delta _\beta
^\alpha $ and the Jacobi matrix $Y\left( t\right) =\left( Y_\beta ^\alpha
\left( t\right) =\frac{\delta U^\alpha }{\delta u^\beta }\left( t,u,w\right)
\right) ,$ which satisfy the matrix equation%
\begin{equation}
Y\left( t\right) =I+\int_0^t\sigma _{\widehat{\alpha }}^{\prime }\left(
U\left( s\right) \right) Y\left( s\right) \circ dw^{\widehat{\alpha }}\left(
s\right) +\int_0^tb^{\prime }\left( U\left( s\right) \right) Y\left(
s\right) ds.\eqnum{4.10}
\end{equation}

As a modification of a process $U\left( t,u,w\right) $ one means a such
process $\widehat{U}\left( t,u,w\right) $ that $P^W\{\widehat{U}\left(
t,u,w\right) =U\left( t,u,w\right) $ for all $t\geq 0\}=1$ a.s.

It is known this result for flows of diffeomorphisms of flat spaces
[24-26,7]:

{\bf Theorem 4.2. }{\it Let }$U\left( t,u,w\right) $ {\it be the solution of
the equation (4.8) (or (4.7)) on Wiener space }$\left( W_0^r,P^W\right) $ .%
{\it Then we can choose a modification $\widehat{U}\left( t,u,w\right) $of
this solution when the map }$u\rightarrow U\left( t,u,w\right) $\text{{\it \
is a diffeomorphism }}${\cal R}^{n+m}$ {\it a.s. for every }$t\in [0,\infty
).$

Process $u=\widehat{U}\left( t,u,w\right) $ is constructed by using
equations
$$
dU_t^\alpha =\sigma _{\widehat{\alpha }}^\alpha \left( U_t\right) \circ
\delta w^{\widehat{\alpha }}\left( t\right) -b^\alpha \left( U_t\right) dt,
$$
$$
U_0=u.
$$
Then for every fixed $T>0$ we have
$$
U\left( T-t,u,w\right) =\widehat{U}\left( t,U\left( T,u,w\right) ,\widehat{w}%
\right)
$$
for every 0$\leq t\leq T$ and $u$ $P^W$-a.s., where the Wiener process $
\widehat{w}$ is defined as $\widehat{w}\left( t\right) =w\left( T-t\right)
-w\left( T\right) ,0\leq t\leq T.$

Now we can extend the results on flows of diffeomorphisms of stochastic
processes to v-bundles. The solution $U\left( t,u,w\right) $ of the equation
(3.1) can be considered as the set of maps $U_t:u\rightarrow U\left(
t,u,w\right) $ from ${\cal E_N}$ to $\hat {{\cal }}E_N= {\cal E_N \cup
\{\bigtriangleup \}.}$

{\bf Theorem 4.3. }{\it A process }$|U|\left( t,u,w\right) $\ {\it has such
a modification, for simplicity let denote it also as } $U\left( t,u,w\right)
,$ {\it that the map }$U_t(w):u\rightarrow U\left( t,u,w\right) $ {\it %
belongs to the class }$C^\infty $ {\it for every }$f\in F_0\left( {\cal E_N}%
\right) $ {\it and all fixed }$t\in [0,\infty )$ {\it a.s. In addition, for
every }$u\in U$ {\it and }$t\in [0,\infty )$ {\it the differential of map }$%
u\rightarrow U\left( t,u,w\right) ,$

$$
U\left( t,u,w\right) _{*}:T_{u\,}\left( U\left( t,u,w\right) \right)
\rightarrow T_{U\left( t,u,w\right) }\left( {\cal E_N}\right) ,
$$
{\it is an isomorphism, a.s., in the set }$\{w:U\left( t,u,w\right) \in
{\cal E_N\ \}.}$

{\rm Proof. }Let $u_0\in ${\it \ }${\cal E_N}$ and fix $t\in [0,\infty )$ .
We can find a sequence of coordinate carts $U_1,U_2,...,U_p\subset {\cal %
E_N\ }$ that for almost all $w$ that $U\left( t,u_0\subset w\right) \in
{\cal E_N}$ there is an integer $p>0$ that $\{U\left( s,u_0,w\right) :s\in
[\left( k-1\right) t/p,kt/p]\}\subset {\cal U_k\ ,}\left( k=1,2,...p\right)
. $ According to the theorem 4.2 we can conclude that for every coordinate
cart ${\cal U\ }$ and $\{U\left( s,u_0,w\right) ;s\in [0,1]\}\subset {\cal %
U\ }$ a map $v\rightarrow U\left( t_0,v,w\right) $ is a diffeomorphism in
the neighborhood of $v_{0.}$The proof of the theorem follows from the
relation $U\left( t,w_0,w\right) =[U_{t/p}\left( \theta _{\left( p-1\right)
t/p}w\right) \circ ...\circ U_{t/p}\left( \theta _{t/p}w\right) \circ
U_{t/p}]\left( u_0\right) ,$ where $\theta _t:W_0^r\rightarrow W_0^r$ is
defined as $\left( \theta _tw\right) \left( s\right) =w\left( t+s\right)
-w\left( t\right) .\diamondsuit $

Let $A_0,A_1,...,A_r\in {\bf X\left( {\cal E_N}\right) }$ and $U_t=\left(
U\left( t,u,w\right) \right) $ is a flow of diffeomorphisms on ${\bf {\cal %
E_N}}$ . Then $\widetilde{A}_0,\widetilde{A}_1,...,\widetilde{A}_r\in {\bf X}%
\left( GL\left( {\bf {\cal E_N}}\right) \right) $ define a flow of
diffeomorphisms $r_t=\left( r\left( t,r,w\right) \right) $ on $GL\left( {\bf
{\cal E_N}}\right) $ with $\left( r\left( t,r,w\right) \right) =\left(
U\left( t,u,w\right) ,e\left( t,u,w\right) \right) ,$ where $r=\left(
u,e\right) $ and $e\left( t,r,u\right) =U\left( t,u,w\right) _{*}e$ is the
differential of the map $u\rightarrow U\left( t,u,w\right) $ satisfying the
property $U\left( t,u,w\right) _{*}e=[U\left( t,u,w\right) _{*}e_0,U\left(
t,u,w\right) _{*}e_1,...,U\left( t,u,w\right) _{*}e_{q-1}].$ In local
coordinates $A_{\widehat{\alpha }}\left( u\right) =\sigma _{\widehat{\alpha }%
}^\alpha \delta _\alpha ,\left( \alpha =1,2,...,r\right) ,A_0\left( u\right)
=b^\alpha \left( u\right) \delta _\alpha ,e_{\underline{\beta }}^\alpha
\left( t,u,w\right) =Y_{\underline{\gamma }}^\alpha \left( t,u,w\right) e_{
\underline{\beta }}^{\underline{\gamma }},$ where $Y_{\underline{\gamma }%
}^\alpha \left( t,u,w\right) $ is defined from (4.10). So we can construct
flows of diffeomorphisms of the bundle ${\bf {\cal E_N}}$ .

\section{Nondegenerate Diffusion in Bundle Spaces}

Let a v-bundle ${\bf {\cal E_N\ }}$ be provided with a positively defined
metric of type (2.10) being compatible with a d- connection $D=\{\Gamma
_{\beta \gamma }^\alpha \}.$ The connection $D$ allows us to roll ${\bf
{\cal E_N\ }}$ along a curve $\gamma \left( t\right) $ $\subset {\cal R}^{n+m%
}$ in order to draw the curve $c\left( t\right) $ on ${\bf {\cal E_N}}$ as
the trace of $\gamma \left( t\right) .\,$ More exactly, let $\gamma
:[0,\infty )\ni t\rightarrow \gamma \left( t\right) \subset {\cal R}^{n+m}\ $
be a smooth curve in ${\cal R}^{n+m},\ r=\left( u,e\right) \in O\left( {\bf
{\cal E_N}}\right) .$ We define a curve $\widetilde{c}\left( t\right)
=\left( c\left( t\right) ,e\left( t\right) \right) $ in $O\left( {\bf {\cal %
E_N}}\right) $ by using the equalities%
$$
\frac{dc^\alpha \left( t\right) }{dt}=e_{\underline{\alpha }}^\alpha \left(
t\right) \frac{d\gamma ^{\underline{\alpha }}}{dt},
$$
\begin{equation}
\frac{de_{\underline{\alpha }}^\alpha \left( t\right) }{dt}=-\Gamma _{\beta
\gamma }^\alpha \left( c\left( t\right) \right) e_{\underline{\alpha }%
}^\gamma \left( t\right) \frac{dc^\beta }{dt}, \eqnum{5.1}
\end{equation}
$$
c^\alpha \left( 0\right) =u^\alpha ,e_{\underline{\alpha }}^\alpha \left(
0\right) =e_{\underline{\alpha }}^\alpha .
$$
Equations (5.1) can be written as
$$
\frac{d\widetilde{c}\left( t\right) }{dt}=\widetilde{L}_\alpha \left(
\widetilde{c}\left( t\right) \right) d\gamma ^\alpha ,
$$
$$
\widetilde{c}\left( 0\right) =r,
$$
where$\{\widetilde{L}_\alpha \}$ is the system of canonical horizontal
vector fields (see (2.21)). Curve $c\left( t\right) =\pi \left( \widetilde{c}%
\left( t\right) \right) $ on ${\cal E_N\ }$ depends on fixing of the initial
frame $p$ in a point $u;$ this curve is parametrized as $c\left( t\right)
=c\left( t,r,\gamma \right) ,r=r\left( u,e\right) .$

Let $w\left( t\right) =\left( w^{\underline{\alpha }}\left( t\right) \right)
$ is the canonical realization of a n+m-dimensional Wiener process.\ We can
define the random curve $U\left( t\right) \subset $ ${\cal E_N}$ in a
similar manner. Consider $r\left( t\right) =\left( r\left( t,r,w\right)
\right) $ as the solution of stochastic differential equations
\begin{equation}
dr\left( t\right) =\widetilde{L}_{\underline{\alpha }}\left( r\left(
t\right) \right) \circ \delta w^{\underline{\alpha }}\left( t\right) ,
\eqnum{5.2}
\end{equation}
$$
r\left( 0\right) =r,
$$
where $r\left( t,r,w\right) $ is the flow of diffeomorphisms on $O\left(
{\cal E_N\ }\right) $ corresponding to the canonical horizontal vector
fields $\widetilde{L}_1,\widetilde{L}_2,...,\widetilde{L}_{q-1}$ and
vanishing drift field $\widetilde{L}_0=0.$ In local coordinates the
equations (5.2) are written as
$$
dU^\alpha \left( t\right) =e_{\underline{\alpha }}^\alpha \left( t\right)
\circ \delta w^{\underline{\alpha }}\left( t\right) ,
$$
$$
de_{\underline{\alpha }}^\alpha \left( t\right) =-\Gamma _{\beta \gamma
}^\alpha \left( U\left( t\right) \right) e_{\underline{\alpha }}^\gamma
\circ \delta u^\beta ,
$$
where $r\left( t\right) =\left( U^\alpha \left( t\right) ,e_{\underline{%
\alpha }}^\alpha \left( t\right) \right) .\,\,$It is obvious that $r\left(
t\right) =\left( U^\alpha \left( t\right) ,e_{\underline{\alpha }}^\alpha
\left( t\right) \right) \in O\left( {\cal E_N\ }\right) $ if $r\left(
0\right) \in O\left( {\cal E_N\ }\right) $ because $\widetilde{L}_{
\underline{\alpha }}$ are vector fields on $O\left( {\cal E_N\ }\right) .$
The random curve $\{U^\alpha \left( t\right) \}$ on ${\cal E_N\ }$ is
defined as $U\left( t\right) =\pi \left[ r\left( t\right) \right] .$ We
point out that $aw=\left( aw\left( t\right) \right) $ is another
(n+m)-dimensional Wiener process and as a consequence the probability law $%
U\left( \cdot ,r,w\right) $ does not depend on $a\in O\left( n+m\right) .$
It depends only on $u=\pi \left( r\right) .$ This law is denoted as $P_w$
and should be mentioned that it is a Markov process because a similar
property has $r\left( \cdot ,r,w\right) .$

{\bf Remark 4.1.\thinspace }{\it \ We can define }$r\left( t,r,w\right) $
{\it as a flow of diffeomorphisms on }$GL\left( {\cal E_N\ }\right) $ {\it %
for every d-connection on }${\cal E_N.}$ {\it In this case $\pi \left[
r\left( \cdot ,r,w\right) \right] $} {\it does not depend only on }$u=\pi
\left( t\right) $ {\it and in consequence we do not obtain a Markov process
by projecting on } ${\cal E_N.}$ {\it The Markov property of diffusion
processes on }${\cal E_N\ }$ {\it is assumed by the conditions of
compatibility of metric and d-connection (2.11) and of vanishing of torsion.}

Now let us show that a diffusion $\{P_u\}$ on ${\cal E_N}$ can be considered
as an A-diffusion process with the differential operator
\begin{equation}
A=\frac 12\Delta _{{\cal E\ }}+b, \eqnum{5.3}
\end{equation}
where $\Delta _{{\cal E\ }}$ is the Laplace-Beltrami operator on ${\cal E_N,}
$
\begin{equation}
\Delta _{{\cal E\ }}f=G^{\alpha \beta }\overrightarrow{D}_\alpha
\overrightarrow{D}_\beta f=G^{\alpha \beta }\frac{\delta ^2f}{\delta
u^\alpha \delta u^\beta }-\{\frac \alpha {\gamma \beta }\}\frac{\delta f}{%
\delta u^\alpha }, \eqnum{5.4}
\end{equation}
where operator $\overrightarrow{D}_\alpha $ is constructed by using
Christoffel d-symbols (2.15) and $b$ is the vector d-field with components
\begin{equation}
b^\alpha =\frac 12G^{\beta \gamma }\left( \{\frac \alpha {\beta \gamma
}\}-\Gamma _{\beta \gamma }^\alpha \right) \eqnum{5.5}
\end{equation}

{\bf Theorem 5.1.}{\it \ The solution of stochastic differential equation
(5.2) on } $O\left( {\cal E_N\ }\right) $ {\it defines a flow of
diffeomorphisms } $r\left( t\right) =\left( r\left( t,r,w\right) \right) $
{\it on }$O\left( {\cal E_N\ }\right) $ {\it and its projection } $U\left(
t\right) =\pi \left( r\left( t \right) \right) $ {\it defines a diffusion
process on } ${\cal E_N}$ {\it corresponding to the differential operator
(5.3).}

{\rm Proof. } Considering $f\left( r\right) \equiv f\left( u\right) $ for $%
r=\left( u,e\right) $ we obtain%
$$
f\left( U(t)\right) -f\left( U\left( 0\right) \right) =f\left( r\left(
t\right) \right) -f\left( r\left( 0\right) \right) =\int_0^r\left(
\widetilde{L}_{\underline{\alpha }}f\right) \left( r\left( s\right) \right)
\circ \delta w^{\underline{\alpha }}=
$$
$$
\int_0^t\widetilde{L}_{\underline{\alpha }}f\left( r\left( s\right) \right)
\delta w^{\underline{\alpha }}+\frac 12\int_0^t\sum\limits_{\underline{%
\alpha }=0}^{q-1}\widetilde{L}_{\underline{\alpha }}\left( \widetilde{L}_{
\underline{\alpha }}f\right) \left( r\left( s\right) \right) ds.
$$
Let us show that $\frac 12\sum\limits_{\underline{\alpha }=0}^{q-1}
\widetilde{L}_{\underline{\alpha }}\left( \widetilde{L}_{\underline{\alpha }%
}f\right) =Af.$ Really, because the operator (5.3) can be written as
$$
A=\frac 12G^{\alpha \beta }\overrightarrow{D}_\alpha \overrightarrow{D}%
_\beta =\frac 12(G^{\alpha \beta }\frac{\delta ^2}{\delta u^\alpha \delta
u^\beta }-\{\frac \alpha {\gamma \beta }\}\frac \delta {\delta u^\alpha })
$$
and taking into account (2.25) we have
$$
\widetilde{L}_{\underline{\alpha }}\left( \widetilde{L}_{\underline{\alpha }%
}f\right) =\widetilde{L}_{\underline{\alpha }}\left( F_{\nabla f}\right) _{
\underline{\alpha }}=\left( F_{\nabla \nabla f}\right) _{\underline{\alpha }
\underline{\alpha }}=\left( \nabla _\gamma \nabla _\delta f\right) e_{
\underline{\alpha }}^\gamma e_{\underline{\alpha }}^\delta .
$$
Now we can write
$$
\sum\limits_{\underline{\alpha }=0}^{q-1}\widetilde{L}_{\underline{\alpha }%
}\left( \widetilde{L}_{\underline{\alpha }}f\right) =\sum\limits_{\underline{%
\alpha }=0}^{q-1}(\overrightarrow{D}_\alpha \overrightarrow{D}_\beta f\ )e_{
\underline{\alpha }}^\alpha e_{\underline{\beta }}^\beta =G^{\alpha \beta }
\overrightarrow{D}_\alpha \overrightarrow{D}_\beta f\
$$
(see (2.26)), which complete our proof.$\diamondsuit $

{\bf Definition 5.1. }{\it The process }$r\left( t\right) =\left( r\left(
t,r,w\right) \right) $ {\it from the theorem 5.1 is called the horizontal
lift of the A-diffusion }$U\left( t\right) $ {\it on }${\cal E_N.}$

{\bf Proposition 5.1. }{\it For every d-vector field }$b=b^\alpha \left(
u\right) \delta _\alpha $ {\it on }${\cal E_N\ }$ {\it provided with the
canonical d-connection structure there is a d-connection }$D=\{\Gamma
_{\beta \gamma }^\alpha \}$ {\it on }${\cal E_N\ ,}$ {\it compatible with
d-metric }$G_{\alpha \beta },$\text{{\it which satisfies the equality (5.5).}%
}

{\rm Proof. }Let define
\begin{equation}
\Gamma _{\beta \gamma }^\alpha =\{\frac \alpha {\beta \gamma }\}+\frac
2{q-1}\left( \delta _\beta ^\alpha b_\gamma -G_{\beta \gamma }b^\alpha
\right) ,\eqnum{5.6}
\end{equation}
where $b_\alpha =G_{\alpha \beta }b^\beta .$ By straightforward calculations
we can verify that d-connection (5.6) satisfies the metricity conditions%
$$
\delta _\gamma G_{\alpha \beta }-G_{\tau \beta }\Gamma _{\gamma \alpha
}^\tau -G_{\alpha \tau }\Gamma _{\gamma \beta }^\tau =0
$$
and that
$$
\frac 12G^{\alpha \beta }\left( \{\frac \gamma {\alpha \beta }\}-\Gamma
_{\alpha \beta }^\gamma \right) =b^\gamma .\diamondsuit
$$

We note that a similar proposition is proved in [7] for, respectively,
metric and affine connections on Riemannian and affine connected manifolds:
M. Anastasiei proposed [28] to define Laplace-Beltrami operator (5.4) by
using the canonical d-connection (2.13) in generalized Lagrange spaces.
Taking into account (2.16) and (2.17) and a corresponding redefinition of
components of d-vector fields (5.5), because of the existence of
multiconnection structure on the space $H^{2n}$ , we conclude that we can
equivalently formulate the theory of d-diffusion on $H^{2n}$-space by using
both variants of Christoffel d-symbols and canonical d-connection.

{\bf Definition 5.2. }{\it For }$A=\frac 12\Delta _{{\cal E\ }}$ {\it an
A-diffusion }$U\left( t\right) $ {\it is called a Riemannian motion on }$%
{\cal E_N.}$

Let an A-differential operator on ${\cal E_N}$ is expressed locally as
$$
Af\left( u\right) =\frac 12a^{\alpha \beta }\left( u\right) \frac{\delta ^2f
}{\delta u^\alpha \delta u^\beta }\left( u\right) +b^\alpha \left( u\right)
\frac{\delta f}{\delta u^\alpha }\left( u\right) ,
$$
where $f\in F\left( {\cal E_N}\right) ,$ matrix $a^{\alpha \beta }$ is
symmetric and nonegatively defined . If $a^{\alpha \beta }\left( u\right) $ $%
\xi _\alpha \xi _\beta >0$ for all $u$ and $\xi =\left( \xi _\alpha \right)
\in {\cal R}^q \backslash \{0\},$ than the operator $A$ is nondegenerate and
the corresponding diffusion is called nondegenerate.

By using a vector d-field $b_\alpha $ we can define the 1-form%
$$
\omega _{\left( b\right) }=b_\alpha \left( u\right) \delta u^\alpha ,
$$
where $b=b^\alpha \delta _\alpha $ and $b_\alpha =G_{\alpha \beta }b^\beta $
in local coordinates. According the de Rham-Codaira theorem [27] we can write%
\begin{equation}
\omega _{\left( b\right) }=dF+\widehat{\delta }\beta +\alpha \eqnum{5.7}
\end{equation}
where $F\in F\left( {\cal E_N}\right) ,\beta $ is a 2-form and $\alpha $ is
a harmonic 1-form. The scalar product of p-forms $\Lambda _p\left( {\cal %
E_N\ }\right) $ on ${\cal E_N}$ is introduced as%
$$
(\alpha ,\beta )_B=\int_{{\cal E_N}}<\alpha ,\beta >\delta u,
$$
where%
$$
\alpha =\sum\limits_{\gamma _1<\gamma _2<...<\gamma _p}\alpha _{\gamma
_1\gamma _2...\gamma _p}\delta u^{\gamma _1}\bigwedge \delta u^{\gamma
_2}\bigwedge ...\bigwedge \delta u^{\gamma _p},
$$
$$
\beta =\sum\limits_{\gamma _1<\gamma _2<...<\gamma _p}\beta _{\gamma
_1\gamma _2...\gamma _p}\delta u^{\gamma _1}\bigwedge \delta u^{\gamma
_2}\bigwedge ...\bigwedge \delta u^{\gamma _p},
$$
$$
\beta ^{\gamma _1\gamma _2...\gamma _p}=G^{\gamma _1\tau _1}G^{\gamma _2\tau
_2}...G^{\gamma _p\tau _p}\beta _{\tau _1\tau _2...\tau _p},
$$
$$
<\alpha ,\beta >=\sum\limits_{\gamma _1<\gamma _2<...<\gamma _p}\alpha
_{\gamma _1\gamma _2...\gamma _p}\left( u\right) \beta ^{\gamma _1\gamma
_2...\gamma _p}\left( u\right) ,
$$
$$
\delta u=\sqrt{|\det G_{\alpha \beta }|}\delta u^0\delta u^1...\delta
u^{q-1}.
$$
The operator $\widehat{\delta }:\Lambda _p\left( {\cal E_N}\right)
\rightarrow \Lambda _{p-1}\left( {\cal E_N}\right) $ from (5.7) is defined
by the equality%
$$
\left( d\alpha ,\beta \right) _p=\left( \alpha ,\widehat{\delta }\beta
\right) _{p-1},\alpha \in \Lambda _{p-1}\left( {\cal E_N}\right) ,\beta \in
\Lambda _p\left( {\cal E_N}\right) .
$$
De Rham-Codaira Laplacian $\Box :\Lambda _p\left( {\cal E_N}\right)
\rightarrow \Lambda _p\left( {\cal E_N}\right) $ is defined by the equality%
\begin{equation}
\Box =-\left( d\widehat{\delta }+\widehat{\delta }d\right) .\eqnum{5.8}
\end{equation}

A form $\alpha \in \Lambda _p\left( {\cal E_N}\right) $ is called as
harmonic if $\Box \alpha =0$ . It is known that $\Box \alpha =0$ if and only
if $d\alpha =0$ and $\widehat{\delta }\alpha =0.$For $f\in F\left( {\cal E_N}%
\right) $ and $U\in {\bf X}\left( {\cal E_N}\right) $ we can define the
operators $gradf\in {\bf X}$ $\left( {\cal E_N}\right) $ and $divU\in
F\left( {\cal E_N}\right) $ by using correspondingly the equalities%
\begin{equation}
gradf=G^{\alpha \beta }\delta _\alpha \delta _\beta f \eqnum{5.9}
\end{equation}
and
\begin{equation}
divU=-\widehat{\delta }\omega _U=\frac 1{\sqrt{|\det G|}}\delta _\alpha
\left( U^\alpha \sqrt{|\det G|}\right) . \eqnum{5.10}
\end{equation}

The Laplace-Beltrami operator (5.3) can be also written as
\begin{equation}
\bigtriangleup _{{\cal E\ }}f=div(gradf)=-\widehat{\delta }\widehat{\delta }%
f \eqnum{5.11}
\end{equation}
for $F\left( M\right) .$

Let suggest that ${\cal E_N}$ is compact and oriented and $\{P_u\}$ be the
system of diffusion measures defined by a A-operator (5.3). Because ${\cal %
E_N}$ is compact $P_{u\text{ }}$ is the probability measure on the set $
\widehat{W}\left( {\cal E_N}\right) =W\left( {\cal E_N}\right) $ of all
continuous paths in ${\cal E_N}$ .

{\bf Definition 5.3. }{\it The transition semigroup }$T_t$ {\it of
A-diffusion is defined by the equality%
$$
\left( T_tf\right) \left( u\right) =\int\limits_{W({\cal E_N)\ }}f\left(
w\left( t\right) \right) P_u\left( dw\right) ,f\in C\left( {\cal E_N}\right)
.
$$
}

For a connected open region $\Omega \subset {\cal E_N\ }$ we define $\rho
^\Omega w\in \widehat{W}\left( \Omega \right) ,w\in \widehat{W}\left( {\cal %
E_N}\right) $ by the equality%
$$
(\rho ^\Omega w)\left( t\right) =\langle _{\Delta ,{\em if\ }t\geq \tau
_\Omega \left( w\right) ,}^{w(t),{\em if\ }t<\tau _\Omega \left( w\right) ,}
$$
where $\tau _\Omega \left( w\right) =\inf \{t:w\left( t\right) \notin \Omega
\}.$ We denote the image-measure $P_u\left( u\in \Omega \right) $ on map $%
\rho ^\Omega $ as $P_u^\Omega $ ; this way we define a probability measure
on $\widehat{W}\left( \Omega \right) $ which will be called as the minimal
A-diffusion on $\Omega .$ The transition group of this diffusion is
introduced as
$$
\left( T_t^\Omega f\right) \left( u\right) =\int\limits_{W\left( \Omega
\right) }f\left( w\left( t\right) \right) P_u^\Omega \left( dw\right) =
$$
$$
\int\limits_{W\left( {\cal E_N\ }\right) }f\left( w\left( t\right) \right)
I_{\{\tau _\Omega \left( w\right) >t\}}P_u\left( dw\right) ,f\in C_p\left(
\Omega \right) .
$$

{\bf Definition 5.4. }{\it The Borel measure $\mu \left( du\right) $} {\it %
on }${\cal E_N}$ {\it is called an invariant measure on A-diffusion }$\{P_u\}%
{\it \ }$\text{{\it if $\int\limits_{{\cal E_N}}$}}$T_tf\left( u\right) \mu
\left( du\right) =$\text{{\it $\int\limits_{{\cal E_N}}$}}$f\left( u\right)
\mu \left( du\right) $ {\it for all }$f\in C\left( {\cal E_N}\right) .$

{\bf Definition 5.5. }{\it An A-diffusion }$\{P_u\}$ {\it is called
symmetrizable (locally symmetrizable) if there is a Borel measure $\nu
\left( du\right) $} {\it on }${\cal E_N}$ $\left( \nu ^\Omega \left(
du\right) \text{ on }\Omega \right) {\it \ }$\text{{\it that }}

\begin{equation}
\int\limits_{{\it \ }{\cal E_N\ }}T_tf\left( u\right) g\left( u\right) \nu
\left( du\right) =\int\limits_{{\it \ }{\cal E_N\ }}f\left( u\right)
T_tg\left( u\right) \nu \left( du\right) \eqnum{5.12}
\end{equation}
{\it for all }$f,g\in C\left( {\cal E_N}\right) $ {\it and }

$$
(\int\limits_\Omega T_t^\Omega f\left( u\right) g\left( u\right) \nu ^\Omega
\left( du\right) =\int\limits_\Omega f\left( u\right) T_t^\Omega g\left(
u\right) \nu ^\Omega \left( du\right)
$$
{\it for all }$f,g\in C\left( \Omega \right) ).$

The fundamental properties of A-diffusion measures are satisfied by the
following theorem and corollary:

{\bf Theorem 5.2. }{\it a) An A-diffusion is symmetrizable if and only if $
\widehat{\delta }\beta =\alpha =0$} {\it (see (5.8)); this condition is
equivalent to the condition that }$b=gradF,F\in F\left( {\cal E_N}\right) $
{\it and in this case the invariant measures are of type }$C\exp [2F\left(
u\right) ]du,$ {\it where }$C=const.$

{\it b) An A-diffusion is locally symmetrizable if and only if $\widehat{%
\delta }\beta =0$} {\it (see (5.8)) or, equivalently , }$dw_{\left( b\right)
}=0.$

{\it c) A measure }$cdu$ {\it (constant }$c>0)$ {\it is an invariant measure
of an A-diffusion if and only if }$dF=0$ {\it (see (5.8)) or, equivalently, $
\widehat{\delta }w_{\left( b\right) }=-divb=0.$}

{\bf Corollary 5.1. }{\it An A-diffusion is symmetric with respect to a
Riemannian volume }$du$ {\it (i.e. is symmetrizable and the measure $\nu $}
{\it in (5.14) coincides with }$du)$ {\it if and only if it is a Brownian
motion on }${\cal E_N.}$

We omit the proofs of the theorem 5.2 and corollary 5.1 because they are
similar to those presented in [7] for Riemannian manifolds. In our case we
have to change differential forms and measures on Riemannian spaces into
similar objects on ${\cal E_N}$

\section{Heat Equations for Distinguished Tensor Fields in Vector Bundles}

To generalize the results presented in Section IV to the case of d-tensor
fields in ${\cal E_{N\text{ }}\ }$we use the Ito idea of stochastic parallel
transport [27,30] (correspondingly adapted to transports in vector bundles
provided with N-connection structure).

\subsection{Scalarized Tensor d-fields and Heat Equations}

Consider a compact bundle ${\cal E_{N\text{ \thinspace \thinspace }}}$and
the bundle of orthonormalized adapted frames on ${\cal E_N}$ denoted as $%
O\left( {\cal E_N}\right) .$ Let $\{\widetilde{L}_0,\widetilde{L}_1,...,
\widetilde{L}_{q-1}\}$ be the system of canonical horizontal vector fields
on $O\left( {\cal E_N}\right) $ (with respect to canonical d-connection $
\overrightarrow{\Gamma }_{\beta \gamma }^\alpha .$ The flow of
diffeomorphisms $r\left( t\right) =r\left( t,r,w\right) $ on $O\left( {\cal %
E_N}\right) $ is defined through the solution of equations
$$
dr\left( t\right) =\widetilde{L}_{\underline{\alpha }}\left( r\left(
t\right) \right) \circ \delta w^{\underline{\alpha }}\left( t\right) ,
$$
$$
r\left( 0\right) =r,
$$
and this flow defines a diffusion process, the horizontal Brownian motion on
$O\left( {\cal E_N}\right) ,$ which corresponds to the differential operator
\begin{equation}
\frac 12\Delta _{O\left( {\cal E_N}\right) }=\frac 12\sum\limits_{\underline{%
\alpha }}\widetilde{L}_{\underline{\alpha }}\left( \widetilde{L}_{\underline{%
\alpha }}\right) . \eqnum{6.1}
\end{equation}
For a tensor d-field $S\left( u\right) =S_{\beta _1\beta _2...\beta
_q}^{\alpha _1\alpha _2...\alpha _p}\left( u\right) $ we can define its
scalarization $F_S\left( r\right) =F_{S\underline{\beta }_1\underline{\beta }%
_2...\underline{\beta }_q}^{\underline{\alpha }_1\underline{\alpha }_2...
\underline{\alpha }_p}$ (a system of smooth functions on $O\left( {\cal E_N}%
\right) $) similarly as we have done in Section II, but in our case by using
frames satisfying conditions (2.23) in order to deal with bundle $O\left(
{\cal E_N}\right) .$

The action of Laplace-Beltramy operator on d-tensor fields is defined as%
$$
\left( \Delta T\right) _{\beta _1\beta _2...\beta _q}^{\alpha _1\alpha
_2...\alpha _p}=G^{\alpha \beta }\left( \overrightarrow{D}_\alpha \left(
\overrightarrow{D}_\beta T\right) \right) _{\beta _1\beta _2...\beta
_q}^{\alpha _1\alpha _2...\alpha _p}=G^{\alpha \beta }T_{\beta _1\beta
_2...\beta _q;\alpha \beta }^{\alpha _1\alpha _2...\alpha _p},
$$
where $\overrightarrow{D}T$ is the covariant derivation with respect to $
\overrightarrow{\Gamma }_{\beta \gamma }^\alpha .$ We can calculate (by
putting formula (2.21) into (6.1) that
$$
\Delta _{O\left( {\cal E_N}\right) }(F_{S\beta _1\beta _2...\beta
_q}^{\alpha _1\alpha _2...\alpha _p})=\left( F_{\Delta S}\right) _{\beta
_1\beta _2...\beta _q}^{\alpha _1\alpha _2...\alpha _p}.
$$

For a given d-tensor field $S=S\left( u\right) $ let defined this system of
functions on $[0,\infty )\times O\left( {\cal E_N}\right) $:

$$
V_{\beta _1\beta _2...\beta _q}^{\alpha _1\alpha _2...\alpha _p}\left(
t,r\right) =E\left[ F_{S\beta _1\beta _2...\beta _q}^{\alpha _1\alpha
_2...\alpha _p}\left( r\left( t,r,w\right) \right) \right] .
$$

According to the theorem 4.1 $V_{\beta _1\beta _2...\beta _q}^{\alpha
_1\alpha _2...\alpha _p}$ is a unique solution of heat equation
\begin{equation}
\frac{\partial V}{\partial t}=\frac 12\Delta _{O\left( {\cal E_N}\right)
}V, \eqnum{6.2}
\end{equation}
$$
V_{|t=0}=F_{S\beta _1\beta _2...\beta _q}^{\alpha _1\alpha _2...\alpha _p}.
$$

In a similar manner we can construct unique solutions of heat equations
(6.2) for the case when instead of differential forms one considers ${\cal %
R}^{m+n}$-tensors (see [7] for details concerning Riemannian manifolds).\ We
have to take into account the torsion components of the canonical
d-connection on ${\cal E_N}$ .

\subsection{Boundary Conditions}

We analyze the heat equations for differential forms on a bounded space $%
{\cal E_N:}$

\begin{equation}
\frac{\partial \alpha }{\partial t}=\frac 12\Box \alpha ,\eqnum{6.3}
\end{equation}
$$
\alpha _{|t=0}=f,
$$
\begin{equation}
\alpha _{norm}=0,(d\alpha )_{norm}=0, \eqnum{6.4}
\end{equation}
where $\Box $ is the de Rham-Codaira Laplacian (5.8),%
$$
\alpha _{norm}=\theta _{q-1}\left( u\right) du^{q-1},
$$
$$
\left( d\alpha \right) _{norm}=\sum\limits_{\gamma =1}^q\left( \frac{\delta
\alpha }{\delta u^{q-1}}-\frac{\delta \alpha _{q-1}}{\delta u^\gamma }%
\right) \delta u^{q-1}\bigwedge \delta u^\gamma .0
$$
We consider the boundary of ${\cal E_N\ }$to be a manifold of dimension $%
q=m+n$ and denote by $\overbrace{{\cal E_N}}$ the interior part of ${\cal E_N%
}$ and as $\partial {\cal E_N}$ the boundary of ${\cal E_N.}$ In the
vicinity $\overbrace{{\cal U}}$ of the boundary we introduce the system of
local coordinates $u=\{\left( u^\alpha \right) ,u^{q-1}\geq 0\}$ for every $%
u\in {\cal U\ }$and $u\in {\cal U\ }\cap \partial {\cal E_N}$ if and only if
$u^{q-1}=0.$

The scalarization of 1-form $\alpha $ is defined as%
$$
\left[ F_\alpha \right] _{\underline{\beta }}\left( r\right) =\theta _\beta
\left( u\right) e_{\underline{\beta }}^\beta ,r=\left( u^\beta ,e_{
\underline{\beta }}^\beta \right) \in {\cal O\left( E_N\right) .}
$$
Conditions (6.4) are satisfied if and only if
$$
e_{q-1}^\alpha \left[ F_\alpha \right] _{\underline{\alpha }}\left( r\right)
=0
$$
and
$$
e_\beta ^{\underline{\beta }}\frac \delta {\delta u^{q-1}}\left[ F_\alpha
\right] _{\underline{\beta }}\left( r\right) =0,
$$
\underline{$\alpha $}=0,1,2,...,q-1, where $e_\beta ^{\underline{\alpha }}$
is inverse to $e_{\underline{\beta }.}^\alpha $

Now we can formulate the Cauchy problem for differential 1-forms (6.3) and
(6.4) as a corresponding problem for ${\cal R}^{n+m}$-valued equivariant
functions $V_{\underline{\alpha }}\left( t,r\right) $ on ${\cal O\left(
E_N\right) :}$

\begin{equation}
\frac{\partial V_{\underline{\alpha }}}{\partial r}\left( t,r\right) =\frac
12\{\Delta _{{\cal O\left( E_N\right) \ }}V_{\underline{\alpha }}\left(
t,r\right) +R_{\underline{\alpha }}^{\underline{\beta }}\left( r\right) V_{
\underline{\beta }}\left( t,r\right) , \eqnum{6.5}
\end{equation}
$$
V_{\underline{\alpha }}\left( 0,r\right) =\left( F_f\right) _{\underline{%
\alpha }}\left( r\right) ,(\beta =0,1,...,q-2),(\underline{\alpha },
\underline{\beta }=0,1,...,q-1),
$$
$$
e_\beta ^{\underline{\beta }}\frac \delta {\delta u^{q-1}}V_{\underline{%
\beta }}\left( t,r\right) _{|\partial {\cal O\left( E_N\right) }}=0,f_{q-1}^{
\underline{\beta }}V_{\underline{\beta }}\left( t,r\right) _{|\partial {\cal %
O\left( E_N\right) \ }}=0,
$$
where $R_{\underline{\alpha }}^{\underline{\beta }}\left( r\right) $ is the
scalarization of the Ricci d-tensor and $\partial {\cal O\left( E_N\right) =}
$ $\{r=(u,e)\in {\cal O\left( E_N\right) ,}$ $u\in \partial {\cal E_N}$ \}.

The Cauchy problem (6.5) can be solved by using the stochastic differential
equations for the process $\left( U\left( t\right) ,c\left( t\right) \right)
$ on ${\cal R}_{+}^{n+m}\times {\cal R}^{{( n+m)}^2}:$

$$
dU_t^\alpha =e_{\widehat{\beta }}^\alpha \left( t\right) \circ \delta B^{
\widehat{\beta }}\left( t\right) +\delta _{q-1}^\alpha \delta \varphi \left(
t\right) ,
$$
$$
de_{\underline{\beta }}^\alpha \left( t\right) =-\overrightarrow{\Gamma }%
_{\beta \gamma }^\alpha \left( U\left( t\right) \right) e_{\underline{\beta }%
}^\gamma \left( t\right) \circ \delta U^{\underline{\beta }}\left( t\right)
=
$$
\begin{equation}
-\overrightarrow{\Gamma }_{\beta \gamma }^\alpha \left( U\left( t\right)
\right) e_{\underline{\beta }}^\gamma \left( t\right) e_{\widehat{\tau }%
}^\beta \left( t\right) \circ \delta B^{\widehat{\tau }}\left( t\right) -
\overrightarrow{\Gamma }_{q-1\tau }^\alpha \left( U\left( t\right) \right)
e_{\underline{\beta }}^\tau \left( t\right) \delta \varphi \left( t\right)
, \eqnum{6.6}
\end{equation}
$$
\left( \widehat{\beta },\widehat{\tau }=1,2,...,q-1\right)
$$
where $B^\alpha \left( t\right) $ is a $\left( n+m\right) $-dimensional
Brownian motion, $U\left( t\right) $ is a nondecreasing process which
increase only if $U\left( t\right) \in \partial {\cal E_N}$ . In [7]
(Chapter IV,7) it is proved that for every Borel probability measure $\mu $
on ${\cal R}_{+}^{n+m}\times{\cal R}^{\left( n+m\right) ^2}$ there is a unique
solution $\left( U\left( t\right) ,c\left( t\right) \right) $ of equations
(6.6) with initial distribution $\mu .$ Because if%
$$
G_{\alpha \beta }\left( U\left( 0\right) \right) e_{\underline{\alpha }%
}^\alpha \left( 0\right) e_{\underline{\beta }}^\beta \left( 0\right)
=\delta _{\underline{\alpha }\underline{\beta }}
$$
then for every $t\geq 0$
$$
G_{\alpha \beta }\left( U\left( t\right) \right) e_{\underline{\alpha }%
}^\alpha \left( t\right) e_{\underline{\beta }}^\beta \left( t\right)
=\delta _{\underline{\alpha }\underline{\beta }{\em \ }}a.s.
$$
(this is a consequence of the metric compatibility criterions (2.10)) we
obtain a diffusion process $r\left( t\right) =\left( U\left( t\right)
,c\left( t\right) \right) $ on ${\cal O\left( E_N\right) .}$ This process is
called the horizontal Brownian motion on the bundle ${\cal O\left(
E_N\right) }$ with a reflecting bound . Let introduce the canonical
horizontal fields (as in (2.21))%
$$
\left( \widetilde{L}_{\underline{\alpha }}F\right) \left( r\right) =e_{
\underline{\alpha }}^\alpha \frac{\partial F\left( r\right) }{\partial
u^\alpha }-\overrightarrow{\Gamma }_{\beta \gamma }^\alpha \left( u\right)
e_{\underline{\alpha }}^\gamma e_{\underline{\tau }}^\beta \frac{\partial
F\left( r\right) }{\partial e_{\underline{\tau }}^\alpha },r=\left(
u,e\right) ,
$$
define the Bochner Laplacian as%
$$
\Delta _{{\cal O\left( E_N\right) \ }}=\sum\limits_{\underline{\alpha }=1}^q
\widetilde{L}_{\underline{\alpha }}\left( \widetilde{L}_{\underline{\alpha }%
}\right)
$$
and put%
$$
\alpha ^{q-1q-1}\left( r\right) =G^{q-1q-1}(u),\alpha _{\underline{\beta }%
}^{q-1\beta }=-e_{\underline{\beta }}^\tau \overrightarrow{\Gamma }_{q-1\tau
}^\beta \left( u\right) G^{q-1q-1}.
$$

{\bf Theorem 6.1.}{\it \ Let }$r\left( t\right) =\left( U\left( t\right)
,c\left( t\right) \right) $ {\it be a horizontal Brownian motion with
reflecting bound giving as a solution of equations (6.6). Then for every
smooth function }$\;S\left( t,r\right) $ {\it on }$[0,\infty )\times $ $%
{\cal O\left( E_N\right) \ }$ {\it we have }
$$
dS\left( t,r(t)\right) =\widetilde{L}_{\widehat{\alpha }}S\left( t,r\left(
t\right) \right) \delta B^{\widehat{\alpha }}+
$$
$$
\{\frac 12\left( \Delta _{{\cal O\left( E_N\right) }}S\right) \left(
t,r\left( t\right) \right) +\frac{\partial S}{\partial t}\left( t,r\left(
t\right) \right) \}dt+\left( \widetilde{U}_{q-1}S\right) \left( t,r\left(
t\right) \right) \delta \varphi \left( t\right) ,
$$
{\it where $\widetilde{U}_{q-1}$} {\it is the horizontal lift of the vector
field }$U_{q-1}=\frac \delta {\delta u^{q-1}}$ {\it defined as
$$
\left( \widetilde{U}_{q-1}S\right) \left( t,r\right) =\frac{\delta S}{\delta
u^{q-1}}\left( t,r\right) +\frac{\alpha _{\underline{\beta }}^{q-1\beta }}{%
\alpha ^{q-1q-1}\left( r\right) }\frac{\partial S}{\partial e_{\underline{%
\beta }}^\beta }\left( t,r\right)
$$
and
$$
\delta U^{q-1}\left( t\right) \delta U^{q-1}\left( t\right) =\alpha
^{q-1q-1}\left( r\left( t\right) \right) dt,\delta U^{q-1}\left( t\right)
\delta e_{\underline{\beta }}^\beta \left( t\right) =\alpha _{\underline{%
\beta }}^{q-1\beta }\left( r\left( t\right) \right) dt.
$$
}

The proof of this theorem is a straightforward consequence of the Ito
formula (see (a6) in the Appendix) and of the property that $\sum_{
\underline{\alpha }}e_{\underline{\alpha }}^\alpha \left( t\right) e_{
\underline{\alpha }}^\beta \left( t\right) =G^{\alpha \beta }\left( U\left(
t\right) \right) \,$ (see (2.26)).

Finally, in this subsection, we point out that for diffusion processes we
are also dealing with the so-called (A,L)-diffusion for bounded manifolds
(see, for example, [7] and formula (a12)) which is defined by second order
operators $A$ and $L$ given correspondingly on ${\cal E_N}$ and $\partial
{\cal E_N}$ .

\section{Discussion}

In the present paper we have given a geometric evidence for a generalization
of stochastic calculus on spaces with local anisotropy . It was possible a
consideration rather similar to that for Riemannian manifolds by using
adapted to nonlinear connection lifts to tangent bundles [31] and
restricting our analysis to the case of v-bundles provided with compatible
N-connection, d-connection and metric structures. We emphasize that in the
so-called almost Hermitian model of generalized Lagrange geometry [11,12]
this condition is naturally satisfied. As a matter of principle we can
construct diffusion processes on every space ${\cal E_N}$ provided with
arbitrary d-connection structure. In this case we can formulate all results
with respect to an auxiliary convenient d-connection, for instance, induced
by the Christoffel d-symbols (2.15), and then by using deformations of type
(2.16) (or (2.17)) we shall find the deformed analogous of stochastic
differential equations and theirs solutions.

When the results of this paper have been communicated during the Iasi
Academic Days, Romania, October 1994 [32] Academician R.\ Miron and
Professor M. Anastasiei pointed our attention to the pioneer works on the
theory of diffusion on Finsler manifolds with applications in biology by
P.L. Antonelli and T.J.Zastavniak [1,33]. Here we remark that because on
Finsler spaces the metric in general is not compatible with connection the
definition of stochastic processes is very sophisticated. Perhaps, the
 uncompatible metric and connection structures are more convenient for modeling
of stochastic processes in biology and this is successfully exploited by the
mentioned authors in spite of the fact that in general it is still unclear
the possibility and manner of definition of metric relations in biology. As
for formulation of physical models of diffusion in anisotropic media and on
locally anisotropic spaces we have to pay a due attention to the mutual
concordance of the laws of transport (i.e. of connections) and of metric
properties of the space, which in physics plays a crucial role. This allows
us to define the Laplace-Beltrami, gradient and divergence operators and in
consequence to give the mathematical definition of diffusion process on
la-spaces in a standard manner.

\acknowledgments

The author would like to acknowledge the essential discussions with
Academician R. Miron and Professor M. Anastasiei on the subject of the paper.

\appendix

\section{Stochastic Equations and Diffusion Processes on Euclidean Spaces}

We summarize the results necessary for considerations in the original part
of this paper. Details on stochastic calculus and diffusion can be found,
for example, in works [7,18-23].

\subsection{Basic Concepts and Notations}

Let consider the probability space $\left( \Omega ,{\cal F}, P\right)
,$ where $\left( \Omega ,{\cal F}\right) $ is a measurable space, called the
sample space, on which a probability measure $P$ can be placed. A stochastic
process is a collection of random variables $X=\{X_L;0\leq t<\infty \}$ on $%
\left( \Omega ,{\cal F}\right) ,$ which take values in a second measurable
space $\left( S,{\cal B\ }\right) ,\,$ called the state space. For our
purposes we suggest that $\left( S,{\cal B\ }\right) $ is locally a
r-dimensional Euclidean space equipped with a $\sigma $-field of Borel sets,
when we have the isomorphism $S\cong {\cal R}^r$ and ${\cal B}\cong
B\left( {\cal R}^r\right) $ , where ${\cal B\left( U\right) }$ denotes the
smallest $\sigma $-field containing all open sets of a topological space $%
{\cal U.}$ The index $t\in [0,\infty )$ of the random variables $X_t$ will
admit a convenient interpretation as time.

We equip the sample space$\left( \Omega ,{\cal F}\right) $ with a
filtration, i.e. we consider a nondecreasing family ${\cal \{F,}$ $t\geq 0\}$
of sub $\sigma $-fields of ${\cal F}:{\cal F}_s \subseteq {\cal F}_t \subseteq {\cal F} $
for 0$\leq s<t<\infty .$ We set ${\cal F}_\infty =$ $\sigma \left(
\bigcup\limits_{t\geq 0}{\cal F}\right) .$

One says that a sequence $X_n$ converges almost surely (one writes in brief
a.s.) to $X$ if for all $\omega \in \Omega ,$ excepting subsets of zero
probability, one holds the convergence%
$$
\lim \limits_{n\rightarrow \infty }X_n\left( \omega \right) =X\left( \omega
\right) .
$$

A random variable $X_t$ is called p-integrable if%
\begin{equation}
\int\limits_\Omega \left| X\left( \omega \right) \right| ^pP\left( d\omega
\right) <\infty ,p>0,\omega \in \Omega ,a.s. \eqnum{a1}
\end{equation}
($X_{t\text{ }}$ is integrable if (a1) holds for $p=1).$ For an integrable
variable $X$ the number
$$
E\left( X\right) =\int\limits_\Omega X\left( w\right) P\left( d\omega
\right)
$$
is the mathematical expectation of $X$ with respect to the probability
measure $P$ on $\left( \Omega ,{\cal F}\right) .$

Using a sub- $\sigma $-field ${\cal G}$ of $\sigma $-field ${\em F}$ we can
define the value%
$$
E\left( X,{\cal G}\right) =\int\limits_{{\cal G\ }}X\left( \omega \right)
d\omega
$$
called as the conditional mathematical expectation of $X$ with respect to $%
{\cal G.}$

Smooth random processes are modeled by the set of all smooth functions $%
w:[0,\infty )\ni t\rightarrow w\left( t\right) \in {\cal R}^r ,$ denoted as $%
W^r=C\left( [0,\infty )\rightarrow {\cal R}^r\ \right) .\,$ Set $W^r$ is
complete and separable with respect to the metric
$$
\rho \left( w_1,w_2\right) =\sum\limits_{n=1}^\infty 2^{-n}\left[ \left(
\max \limits_{0\leq t\leq n}\left| w_1\left( t\right) -w_2\left( 2\right)
\right| \right) \bigwedge 1\right] ,
$$
w$_1,w_2\in W^r,$ where $a\bigwedge 1=\min \{a,1\}.$

Let ${\cal B\ }$ ($W^r)$ be a topological $\sigma $-field. As a Borel
cylindrical set we call a set $B\subset W^r,$ defined as $B=\{w:\left( w\left(
t_1\right) ,w\left( t_2\right) ,...,w\left( t_n\right) \right) $ and $%
E\subset {\cal B}\left( {\cal R}^{nr}\right) .$ We define as ${\cal C\subset B\ }$ (%
$W^r)$ the set of all Borel cylindrical sets.

{\bf Definition A1. }{\it A process }$\{X_t,{\cal F}_t,$ $0\leq t<\infty \}$
{\it is said to be a submartingale (or supermartingale ) if for every }$0\leq
s<t<\infty ,$ {\it we have }$\dot P-$a.s. $E\left( X_t|{\cal F}\right) \geq
X_s$ ({\it or }$E\left( X_t|{\cal F}\right) \leq X_s).$ {\it We shall say
that }$\{X_t,{\cal F}_t,$ $0\leq t<\infty \}$ {\it is a martingale if it is
both a submartingale and a supermartingale.}

Let the function $p\left( t,x\right) ,t>0,x\in {\cal R}^r$ is defined as
$$
p\left( t,x\right) =\left( 2\pi t\right) ^{-\frac r2}\exp \left[ -\frac{%
\left| x\right| ^2}{2t}\right]
$$
and $X=\left( X_t\right) _{t\in [0,\infty )}$ is a r-dimensional process
that for all $0<t_1<...<t_m$ and $E_i\in {\cal B}\left( {\cal R}^r \right) ,$ $%
i=1,2,...,m,$

$$
P\{X_{t_1}\in E_1,X_{t_2}\in E_2,...,X_{t_m}\in E_m \}=
$$
\begin{equation}
\int\limits_{{\cal R^r\ }}\mu \left( dx\right) \int\limits_{E_1}p\left(
t_1,x_1-x\right) dx_1\int\limits_{E_2}p\left( t_2-t_1,x_2-x_1\right)
dx_2...\int\limits_{E_m}p\left( t_m-t_{m-1},x_m-x_{m-1}\right) dx_m,
\eqnum{a2} \end{equation}
where $\mu $ is the probability measure on$ \left( {\cal R}^r,B\left(
{\cal R}^r \right) \right) .$

{\bf Definition A2. }{\it A process }$X=\left( X_t\right) $ {\it with the
above stated properties is called a r-dimensional Brownian motion (or a
Wiener process with initial distribution $\mu .$} {\it A probability }$P^X$
{\it on $\left( W^r,{\cal B}\text{ }\left( W^r\right) \right) ,$} {\it where
}$P\{w:w\left( t_1\right) \in E_1,w\left( t_2\right) \in E_2,...,w\left(
t_m\right) \in E_m\}$ {\it is given by the right part of (a.2) is called a
r-dimensional Wiener distribution with initial distribution $\mu .$}

Now, let us suggest that on the probability space $\left( \Omega ,{\cal F,}
\text{ \thinspace }P\right) $  a filtration $ \left( {\cal F}_t \right)
,t\in [0,\infty $ ) is given. We can introduce a r-dimensional
 $ \left( {\cal F}_t\right) $-Brownian motion as a d-dimensional smooth process $X=\left(
X\left( t\right) \right) _{t\in [0,\infty )}, \left( {\cal F}_t \right) $%
-coordinated and satisfying condition
$$
E\left[ \exp \left[ i<\xi ,X_t-X_s>\right] |{\cal F}_s\ \right] =\exp \left[
-\left( t-s\right) \left| \xi \right| ^2/2\right] {\em a.s.\ }
$$
for every $\xi \in {\cal R}^r$ and $0\leq s<t.$

The next step is the definition of the Ito stochastic integral [21,7,18-20].
Let denote as ${\cal L}_2$ the space of all real measurable processes $\Phi
=\{\Phi \left( t,u\right) \}_{t\geq 0}$ on $\Omega ,$ $ \left(
{\cal F}_t\right) $ -adapted for every $T>0,$

$$
\parallel \Phi \parallel _{2,T}^2\doteq E\left[ \int\limits_0^T\Phi ^2\left(
s,\omega \right) ds<\infty \right] ,
$$
where ''$\doteq "$ means ''is defined to be''. For $\Phi \in {\cal L}_2$ we
write%
$$
\parallel \Phi \parallel _2\doteq \sum\limits_{n=1}^\infty 2^{-2}\left(
\parallel \Phi \parallel _{2,n}\bigwedge 1\right) .
$$
We restrict our considerations to processes of type%
\begin{equation}
\Phi \left( t,\omega \right) =f_0\left( \omega \right)
I_{\{t=0\}}+\sum\limits_{i=0}^\infty f_i\left( \omega \right)
I_{(t_it_{i+1}]}\left( t\right) , \eqnum{a3}
\end{equation}
where $I_A\left( B\right) =1,$ if $A\subset B$ and $I_{A\,}\left( B\right)
=0,$ if $A\subseteq B.$

Let denote ${\cal M}_2=$\ \{$X=\left( X_t\right) _{t\geq 0};X$ is a
quadratic integrable martingale on $\left( \Omega ,{\cal F,}\text{ }P\right)
$ referring to $ \left( {\cal F}_t\right) _{t\geq 0}$ and $X_0=0$ a.s. \} and
${\cal M}_2^c=\{ $ $X\in {\cal M}_2;\ $ \thinspace $t\rightarrow X$ is smooth
a.s.\}. For $X\in {\cal M}_2\ $ we use denotations
$$
\left| X\right| _T\doteq E\left[ X_T^2\right] ^{\frac 12},T>0,
$$
and $\left| X\right| =\sum\limits_{n=1}^\infty 2^{-n}\left( \left| X\right|
_n\bigwedge 1\right) .$

Now we can define stochastic integral on $ \left( {\cal F}_t \right) $%
-Brownian motion $B\left( t\right) $ on $\left( \Omega ,{\cal F,}\text{ }%
P\right) $ as a map
$$
{\cal L}_2 {\ni }\Phi \rightarrow I\left( \Phi \right) \in {\cal M}_2^c.%
$$

For a process of type (a3) we postulate
$$
I\left( \Phi \right) \left( t,\omega \right)
=\sum\limits_{i=0}^{n-1}[f_i\left( \omega \right) (B\left( t_{i+1},\omega
\right) -B\left( t_i,\omega \right) )+f_n\left( \omega \right) (B\left(
t,w\right) -B\left( t_n,\omega \right) )]=
$$
\begin{equation}
\sum\limits_{i=0}^\infty f_i\left( B\left( t\bigwedge t_{i+1}\right)
-B\left( t\bigwedge t_i\right) \right) \eqnum{a4}
\end{equation}
for $t_n\leq t\leq t_{n+1},n=0,1,2,....$

Process $I\left( \Phi \right) \in {\cal M}_2^c$ defined by (a4) is called
the stochastic integral of $\Phi \in {\cal L}_2$ on Brownian motion $B\left(
t\right) $ and is denoted as
\begin{equation}
I\left( \Phi \right) \left( t\right) =\int\limits_0^t\Phi \left( s,\omega
\right) dB\left( s,w\right) =\int\limits_0^t\Phi \left( s\right) dB\left(
s\right) . \eqnum{a5}
\end{equation}
It is easy to verify that the integral (a5) satisfies properties:%
$$
\left| I\left( \Phi \right) \right| _T=\parallel \Phi \parallel
_{2,T}=\parallel \Phi \parallel _2,
$$
$$
E\left( I\left( \Phi \right) \left( t\right) ^2\right)
=\sum\limits_{i=0}^\infty E\left[ f_i^2\left( t\bigwedge t_{i+1}-t\bigwedge
t_i\right) \right] =E\left[ \int\limits_0^t\Phi ^2\left( s,w\right)
ds\right]
$$
and
$$
I\left( \alpha \Phi +\beta \Psi \right) \left( t\right) =\alpha I\left( \Phi
\right) \left( t\right) +\beta I\left( \Psi \right) \left( t\right) ,
$$
for every $\Phi ,\Psi \in {\cal L}_2$ $\left( \alpha ,\beta \in {\cal R}%
\right) $ and $t\geq 0.$

Consider a measurable space $\left( \Omega ,{\cal F}\right) $ equipped with
a filtration $\left( {\cal F}_t\ \right) .$ A random time $T$ is a stopping
time of this filtration if the event $\{T\leq t\}$ belongs to the $\sigma $%
-field$\left( {\cal F}_t\ \right) $ for every $t\geq 0.$ A random time is an
optional time of the given filtration if $\{T\leq t\}$ $\in \left( {\cal %
F}_t\ \right) $ for every $t\geq 0.$ A real random process $X=\left(
X_t\right) _{t>0}$ on $\left( \Omega ,{\cal F,}\text{ }P\right) $ is called
a local (${\cal F}_t)$ -martingale if it is adapted to $\left(
{\cal F}_t\ \right) $ and there is a sequence of stopping $\left(
{\cal F}_t\ \right) $-moments $\sigma _n$ with $\sigma _n<\infty ,\sigma
_n\uparrow \infty $ and $X_n=\left( X_n(t)\right) $ is a $\left( {\cal F}_t\ %
\right) $-martingale for every $n=1,2,...,$ where $X_n=X\left( t\bigwedge
\sigma _n\right) .$ If $X_n$ is a quadratic integrable martingale for every $%
n,$ than $X$ is called a local quadratic integrable $\left( {\cal F}_t\ %
\right) $-martingale .

Let denote ${\cal M}_2^{loc}=\{$ $X=\left( X_t\right) _{t\geq 0},X$ is a
locally quadratic integrable $\left( {\cal F}_t\ \right) $-martingale and $%
X_0=0$ a.s.\}\thinspace and ${\cal M}_2^{c,loc}$=\{$X\in {\cal M}_2^{loc}:$ $%
t\rightarrow X_t$ is smooth a.s.\}. In a similar manner with the Brownian
motion we can define stochastic integrals for processes $\Phi \in {\cal L}_2$
and $\Phi \in {\cal L}_2^{loc}$ on $M\subset {\cal M}_2^{loc}\ $(we have to
introduce $M\left( t_j,\omega \right) $ instead of $B\left( t_j,\omega
\right) $ respectively for $t_j=t_{i+1},t_j=t_i,t_i=t_n,t_j=1$ in formulas
(a4)). In this case one denotes the stochastic integral as%
$$
I^M\left( \Phi \right) \left( t\right) =\int\limits_0^t\Phi \left( s\right)
dM\left( s\right) .
$$

A great part of random processes can be expressed as a sum of a mean motion
and fluctuations (Ito processes)
$$
X\left( t\right) =X\left( 0\right) +\int\limits_0^tf\left( s\right)
ds+\int\limits_0^tg\left( s\right) dB\left( s\right) ,
$$
where $\int\limits_0^tf\left( s\right) ds$ defines a mean motion , $%
\int\limits_0^tg\left( s\right) dB\left( s\right) $ defines fluctuations and
$\int dB$ is a stochastic integral on Brownian motion $B\left( t\right) .$
In general such processes are sums of processes with limited variations and
martingales. Here we consider the so-called smooth semimartingales
introduced on a probability space with a given filtration $ \left(
{\cal F}_t\right) _{t\geq 0},$ $M^i\left( t\right) \in {\cal M}_2^{c,loc}$ and $%
A^i\left( t\right) $ being smooth (${\cal F}_t)$-adapted processes with
trajectories having a limited variation and $A^i(0)=0$ (i=1,2,...,r). So a
smooth r-dimensional semimartingale can be written as
\begin{equation}
X^i\left( t\right) =X^i\left( 0\right) +M^i\left( t\right) +A^i\left(
t\right) . \eqnum{a6}
\end{equation}
For processes of type (a6) one holds the Ito formula [21,7,18-20] which
gives us a differential-integral calculus for paths of random processes:%
\begin{equation}
F\left( X\left( t\right) \right) -F\left( X\left( 0\right) \right)
=\int\limits_0^tF_i^{\prime }\left( X\left( s\right) \right) dM^i\left(
s\right) +\frac 12\int\limits_0^tF_{ij}^{\prime \prime }\left( X\left(
s\right) \right) d<M^iM^j>\left( s,\right) , \eqnum{a7}
\end{equation}
where $F_i^{\prime }=\frac{\partial F}{\partial x^i},F_{ij}^{\prime \prime }=
\frac{\partial ^2F}{\partial x^i\partial x^j},<M^iM^j>$ is the quadratic
covariation of processes $M^i$ ,$M^j\in {\cal M}_2,\ $ which really
represents a random process $A=A_t,$ parametrized as a difference of two
natural integrable processes with the property that $M_t N_t - A_t$ is
a $ \left( {\cal F}_t\right) $-martingale. Here we remark that a process $%
Q=Q_t$ is an increasing integrable process if it is $ \left( {\cal F}_t\right)
$-adapted, $Q_0=0,$ the map $t\rightarrow Q_1$ is left-continuous , $%
Q_t\geq 0$ and $E\left( Q_t\right) <\infty $ for every $t\in [0,\infty ).$ A
process $Q$ is called natural if for every bounded martingale $M=\left(
M_t\right) $ and every $t\in [0,\infty )$
$$
E\left[ \int\limits_0^\tau M_sdA_s\right] =E\left[
\int\limits_0^tM_sdA_s\right] .
$$

There is another way of definition of stochastic integration instead of Ito
integral , the so-called Fisk-Stratonovich integral, which obeys the usual
rules of mathematical analysis. Let introduce denotations : ${\cal A}$ is
the set of all such smooth $ \left( {\cal F}_t\right) $-adapted processes $%
A=\left( A_t\right) $ that $A_0=0$ and $t\rightarrow A_t$ is a function with
limited variation on every finite integral a.s.; ${\cal B}$ is the set of
all such $\left( {\cal F}_t\right) $-predictable processes $\Phi =\left(
\Phi _t\right) $ that with the probability one the function $t\rightarrow
\Phi _t$ is a bounded function on every finite interval and $\left( t,\omega
\right) \rightarrow X_t\left( \omega \right) $ is ${\cal C}/B\left({\cal R}^r
 \right) $-measurable; ${\cal O\ }$ is the set of quasimartingales (for
every $X\subset {\cal O\ }$ we have the martingale part $M_X$ and the part
with limited variation ). For every $\Phi \in {\cal B}$ and $X\in {\cal O}$
one defines the scalar product
$$
\left( \Phi \circ X\right) =X\left( 0\right) +\int\limits_0^t\Phi \left(
s,w\right) dM_X\left( s\right) +\int\limits_0^t\Phi \left( s,w\right)
dA_X\left( s\right) ,t\geq 0,
$$
as an element of ${\cal O}$ . One introduces an element $\Phi \circ dX\in d%
{\cal O}$ as
$$
\Phi dX=\Phi \circ dX=d\left( \Phi \circ X\right)
$$
in order to define the symmetric Q-product :%
$$
Y\circ dX=YdX+\frac 12dXdY
$$
for $dX\in d{\cal O}$ and $Y\in {\cal O.\ }$ The stochastic integral $%
\int\limits_0^tY\circ dX$ is called the symmetric Fisk-Stratonovich
integral.

\subsection{Stochastic Differential Equations}

Let denote as ${\cal A}^{c,r}$ the set of functions satisfying the
conditions : $\alpha \left( t,w\right) :[0,\infty )\times W^r\rightarrow
{\cal R}^r\times {\cal R}^r$ are ${\cal B}$ $\left( [0,\infty )\right) \times {\cal %
B}$ $\left( W^r\right) /{\cal B}\left({\cal R}^r\otimes {\cal R}^r\right)$-measurable,
for every $t\in [0,\infty )$ a function $W^r\ni w\rightarrow \alpha \left(
t,w\right) \in {\cal R}^r \times {\cal R}^r$ is
${\cal B}_t \left( W^r\right) /{\cal %
B} \left( {\cal R}^d \otimes {\cal R}^c \right) $-measurable, where
 ${\cal R}^r \times {\cal R}^c$
denotes the set of $r\times c$ matrices, ${\cal B} \left( {\cal R}^r\otimes
{\cal R}^r \right) $ is the topological $\sigma $-field on ${\cal R}^r\times {\cal R}^c$ ,
obtained in the result of identification of ${\cal R}^r\times {\cal R}^c$ with $rc $%
-dimensional Euclidean space.

Suppose that values $\alpha \in {\cal A}^{r,c}$ and $\beta \in {\cal A}^{r,1}
$ are given and consider the next stochastic differential equations for a
r-dimensional smooth process $X=\left( X\left( t\right) \right) _{t\geq 0}:$

\begin{equation}
dX_t^\epsilon =\sum\limits_{j=1}^r\alpha _\gamma ^\epsilon \left( t,X\right)
dB^\gamma \left( t\right) +\beta ^\epsilon \left( t,X\right) dt, \eqnum{a8}
\end{equation}
$$
(\epsilon =1,2,...,r),
$$
for simplicity also written as%
$$
dX_t=\alpha \left( t,X\right) dB\left( t\right) +\beta \left( t,X\right) dt.
$$

As a weak solution (with respect to a c-dimensional Brownian motion $B\left(
t\right) ,B\left( 0\right) =0$ a.s.) of the equations (a8) we mean a
r-dimensional smooth process $X=\left( X\left( t\right) \right) _{t\geq 0},$
defined on the probability space $\left( \Omega ,{\cal F}\text{ ,}P\right) $
with such a filtration of $\sigma $-algebras ${\cal \left( F_t\right)
_{t\geq 0}}$ that $X=X\left( t\right) $ is adapted to ${\cal \left(
F_t\right) _{t\geq 0}}$ , i.e. a map $\omega \in \Omega \rightarrow X(\omega
)\in W^r$ is defined and for every $t\in [0,\infty )$ this map is ${\cal %
F_t/B_t}$ $(W^{r\;})$-measurable; we can define processes $\Phi _\beta
^\delta \left( t,\omega \right) =\alpha _\beta ^\delta \left( t,X\left(
\omega \right) \right) \subset {\cal L}_2^{loc}$ and $\Psi ^\delta \left(
t,\omega \right) =\beta ^\delta \left( t,X\left( \omega \right) \right)
\subset {\cal L}_1^{loc}$; values $X\left( t\right) =\left( X^1\left(
t\right) ,X^2\left( t\right) ,...,X^r\left( t\right) \right) $ and $B\left(
t\right) =\left( B^1\left( t\right) ,B^2\left( t\right) ,...,B^c\left(
t\right) \right) $ satisfy equations
\begin{equation}
X^i\left( t\right) -X^i\left( 0\right) =\sum\limits_{\beta
=1}^c\int\limits_0^t\alpha _\beta ^\delta \left( s,X\right) dB^\beta \left(
s\right) +\int\limits_0^t\beta ^\delta \left( s,X\right) ds, \eqnum{a9}
\end{equation}
with the unit probability, where the integral on $dB^\beta \left( s\right) $
is considered as the Ito integral (a7). The first and the second terms in
(a9) are called correspondingly as the martingale and drift terms.

Let $\sigma \left( t,x\right) =\sigma _j^i\left( t,x\right) $ be a Borel
function $\left( t,x\right) \in [0,\infty )\times {\cal R}^r \rightarrow
{\cal R}^r \otimes {\cal R}^r$ and $b\left( t,x\right) =\{b^i\left( t,x\right) \}$ be a
Borel function $\left( t,x\right) \in [0,\infty )\times {\cal R}^r\rightarrow
{\cal R}^r.$ Then $\alpha \left( t,w\right) =\sigma \left( t,w\left( t\right)
\right) \subset {\cal A}^{r,c}$ and $\beta \left( t,w\right) =b\left(
t,w\left( t\right) \right) \in {\cal A}^{r,1}.$ In this case the stochastic
differential equations (a8) are of the Markov type and can be written in the
form%
\begin{equation}
dX^{i\;}\left( t\right) =\sum\limits_{k=1}^r\sigma _k^i\left( t,X\left(
t\right) \right) dB^k\left( t\right) +b^i\left( t,X\left( t\right) \right)
dt. \eqnum{a10}
\end{equation}
If $\sigma $ and $b$ depend only on $x\in {\cal R^c}$ we obtain a equation
with homogeneous in time $t$ coefficients.

Function $\Phi \left( x,w\right) :{\cal R}^r$ $\times W_0^c\rightarrow
W^r,W_0^c=\{w\in {\cal C}$ $\left( [0,\infty )\rightarrow {\cal R}^r \right)
;w\left( 0\right) =0\}$ is called $\widehat{{\cal E}}\left( {\cal R}^r \times
W_0^c\right) $-measurable if for every Borel probability measure $\mu $ on $%
{\cal R}^r\ $there is a function $\widetilde{\Phi }_\mu \left( x,w\right) :%
{\cal R}^r$ $\times W_0^c\rightarrow W^c,$ which is $\overline{{\cal %
B}({\cal R}^r\times W_0^c)}^{\mu \times P^W}/{\cal B}$ $\left( W^r\right) $%
-measurable for all $x\left( \mu \right) ,\Phi \left( x,w\right) =\widetilde{%
\Phi }_\mu \left( x,w\right) $ and $P^W$-almost all $w$ ($P^W$ is the
c-dimensional Wiener measure on $W_0^c,$ i.e. a distribution $B$ ).

A solution $X=\left( X\left( t\right) \right) $ of the equations (a8) with a
Brownian motion $B=B\left( t\right) $ is called a strong solution if there
is a function $F\left( x,w\right) :{\cal R}^r$ $\times W_0^c\rightarrow W^r,$
which is $\widehat{{\cal E}}\left( {\cal R^r}\times W_0^c\right) $%
-measurable for every $x\in {\cal R}^r$ , $w\rightarrow F\left( x,w\right) $
is $\overline{{\cal B}_t \left( W_0^c\right) }^{P^W}/{\cal B}_t
\left( W^c\right) $-measurable for every $t\geq 0$ and $X=F\left( X\left(
0\right) ,B\right) $ a.s.

We obtain a unique strong solution if for every r-dimensional${\cal \left(
F_t\right) }$-Brownian motion $B=B\left( t\right) \left( B\left( 0\right)
=0\right) $ on the probability space with the filtration ${ \left(
{\cal F}_t\right)} $ and arbitrary $ \left({\cal F}_0\right) $-measurable ${\cal %
R}^r$-valued random vector $X=F\left( \xi ,B\right) $ is a solution of (a8)
on this space with $X\left( 0\right) =\xi $ a.s. So, a strong solution can
be considered as a function $F\left( x,w\right) $ which generates a solution
$X$ of equation (a8) if and only if we shall fix the initial value $%
X\left( 0\right) $ and Brownian motion $B.$

\subsection{Diffusion Processes}

As the diffusion processes one names the class of processes which are
characterized by the Markov property and smooth paths (see details in
[22,23,.7]). \ Here we restrict ourselves with the definition of diffusion
processes generated by second order differential operators on ${\cal R}^r:$

\begin{equation}
Af\left( x\right) =\frac 12\sum\limits_{i,j=1}^ra^{ij}\left( x\right) \frac{%
\partial ^2f}{\partial x^i\partial x^j}\left( x\right)
+\sum\limits_{i=1}^rb^i\left( x\right) \frac{\partial f}{\partial x}\left(
x\right) , \eqnum{a11}
\end{equation}
where $a^{ij}\left( x\right) $ and $b^i\left( x\right) $ are real smooth
functions on ${\cal R}^r$, matrix $a^{ij}\left( x\right) $ is symmetric and
positively defined . Let denote by $\widehat{\cal R}^r={\cal R}^r$ $\cup
\{\Delta \}$ the point compactification of ${\cal R}^r.$ Every function $f$
on ${\cal R}^r$ is considered as a function on $\widehat{\cal R}^r$ with $%
f\left( \Delta \right) =0.\,$ The region of definition of the operator (a11)
is taken the set of doubly differentiable functions with compact carrier,
denoted as $C_K^2\left( {\cal R}^r \right) .$ Let ${\cal B}$ $\left( \widehat{%
W}^r\right) $ be the $\sigma $-field generated by the Borel cylindrical sets,
where $\widehat{W}^r=\{w:[0,\infty )\ni t\rightarrow w\left( t\right) \in $ $
\widehat{\cal R}^r$ is smooth and if $w\left( t\right) =\Delta $ , then $%
w\left( t^{\prime }\right) =\Delta $ for all $t^{\prime }\geq t.$ The value $%
e\left( w\right) =\inf \{t;w\left( t\right) =\Delta ,w\in \widehat{W}^r\}$
is called the explosion time of the path $w.$

{\bf Definition A3. }{\it A system of Markov probability distributions }$%
\{P_x,x\in $ ${\cal R}^r\}$ {\it on $\left( \widehat{W}^r,{\cal B}\left(
\widehat{W}^r\right) \right) ,$} {\it which satisfy conditions : }$%
P_X\{w:w(0)=x\}=1$ {\it for every }$x\in {\cal R}^r;$ $f\left( w\left(
t\right) \right) -f\left( w\left( 0\right) \right) -\int\limits_0^t\left(
Af\right) \left( w\left( s\right) \right) ds$ {\it is a $\left( P_x,{\cal %
B}_t\ (\widehat{W}^r)\right) $}-{\it martingale for every }$f\in C_K^2\left(
{\cal R}^r\ \right) $ {\it and }$x\in {\cal R}^r$ {\it defines a diffusion
measure generated by an operator }$A$ ({\it or }$A$-{\it diffusion).}

{\bf Definition A4. }{\it A random process }$X=\left( X\left( t\right)
\right) $ {\it on }${\cal R}^r$ {\it is said to be a diffusion process,
generated by the operator }$A$ ({\it or simply a }$A$-{\it diffusion
process) if almost all paths } $\left[ t\rightarrow X\left( t\right) \right] $%
{\it \ }$\in \widehat{W}^r$ {\it and probability law of the process }$X$
{\it coincides with }$P_\mu \left( \cdot \right) =\int\limits_{{\it \ }%
{\cal R}^r\ } P_x\left( \cdot \right) \mu \left( dx\right) ,$ {\it where $\mu
$} {\it is the diffusion measure generated by the operator }$A$ {\it and \{}$%
P_x\}$ {\it is the probability law of }$X\left( 0\right) .$

To a given A-diffusion we can associate a corresponding stochastic
differential equation. Let the matrix function $\sigma \left( x\right)
=\left( \sigma _j^i\left( x\right) \right) \in {\cal R}^r \times {\cal R}^r$ defines
$a^{ij}\left( x\right) =\sum\limits_{k=1}^r\sigma _k^i\left( x\right) $ $%
\sigma _k^j\left( x\right) $ and consider the equations%
\begin{equation}
dX^i\left( t\right) =\sum\limits_{k=1}^r\sigma _k^i\left( X\left( t\right)
\right) dB^k\left( t\right) +b^i\left( X\left( t\right) \right) dt. \eqnum{a12}
\end{equation}
There is an ex\-ten\-sion of $\left( \Omega ,{\cal F,}P\right) $ with a
fil\-tra\-tion $\left( {\cal F}_t\right) $
of the prob\-a\-bil\-ity space
$$\left(\widehat{W}^r,{\cal B}\left( \widehat{W}^r\right) ,P_X\right) $$ and
with a filtration ${\cal B}_t\left( \widehat{W}^r\right) $ and a $ %
\left( {\cal F}_t \right) $-Brownian motion $B\left( t\right) $ (see [7] and the
previous subsections in this Appendix) that putting $X\left( t\right)
=w\left( t\right) $ and $e=e\left( w\right) \,$ one obtains for $t\in [0,e)$
$$
X^i\left( t\right) =x^i+\sum\limits_{k=1}^r\int\limits_0^t\sigma _k^i\left(
X\left( s\right) \right) dB^k\left( s\right) +\int\limits_0^tb^i\left(
X\left( s\right) \right) ds.
$$
So $\left( X\left( t\right) ,B\left( t\right) \right) $ is the solution of
the equations (a12) with $X\left( 0\right) =x.$

If bounded regions are considered, diffusion is described by second order
partial differential operators with boundary conditions. Let denote $D={\cal %
R}_t^r=$ $\{x=(x^1,x^2,....x^r);x^r\geq 0$ \}, $\partial D=\{x\in
D,x^r=0\},D^0=\{x\in D;x^r>0\}.$ The Wentzell bound operator is defined as a
map from $C_K^2\left( L\right) $ to the space of smooth functions on $%
\partial D$ of this type:

\begin{equation}
Lf\left( x\right) =\frac 12\sum\limits_{i,j=1}^{r-1}\alpha ^{ij}\left(
x\right) \frac{\partial ^2f}{\partial x^i\partial x^j}\left( x\right)
+\sum\limits_{i=1}^{r-1}\beta ^i\left( x\right) \frac{\partial f}{\partial x}%
\left( x\right) +\mu \left( x\right) \frac{\partial f}{\partial x^r}\left(
x\right) -\rho \left( x\right) Af\left( x\right) , \eqnum{a13}
\end{equation}
where $x\in \partial D,\alpha ^{ij}\left( x\right) ,\beta ^i\left( x\right)
,\mu \left( x\right) $ and $\rho \left( x\right) $ are bounded smooth
functions on $\partial D,\alpha ^{ij}\left( x\right) $ is a symmetric and
nondegenerate matrix, $\mu \left( x\right) \geq 0$ and $\rho \left( x\right)
\geq 0.$

A diffusion process defined by the operators (a11) and (a13) is called a $%
\left( A,L\right) $-diffusion.

\end{document}